\documentclass[useAMS,usenatbib]{mn2e}
\usepackage{multirow,graphicx}

\title[Dust extinction for an unbiased sample of GRB afterglows]{Dust extinctions for an unbiased sample of GRB afterglows}
\author[S. Covino et al.]{S. Covino$^{1}$\thanks{E-mail: stefano.covino@brera.inaf.it}, 
A. Melandri$^{1}$,
R. Salvaterra$^{2}$,
S. Campana$^{1}$, 
S. D. Vergani$^{3,1}$,  \newauthor
M.G. Bernardini$^{1}$,
P. D'Avanzo$^{1}$,
V. D'Elia$^{4,5}$,
D. Fugazza$^{1}$,
G. Ghirlanda$^{1}$,  \newauthor 
G. Ghisellini$^{1}$,   
A. Gomboc$^{6,7}$,
Z.P. Jin$^{8,1}$, 
T. Kruehler$^{9}$,
D. Malesani$^{9}$, \newauthor
L. Nava$^{10}$, 
B. Sbarufatti$^{1}$, 
and G. Tagliaferri$^{1}$\\
$^{1}$INAF / Brera Astronomical Observatory, via Bianchi 46, 23807, Merate (LC), Italy\\
$^{2}$INAF / IASF Milano, via E. Bassini 15, 20133, Milano, Italy\\
$^{3}$GEPI, Observatoire de Paris, CNRS, Univ. Paris Diderot, 5 place Jules Janssen, 92190, Meudon, France\\
$^{4}$INAF / Rome Astronomical Observatory, via Frascati 33, 00040, Monteporzio Catone (Roma), Italy \\
$^{5}$ASI Science Data Centre, Via Galileo Galilei, 00044, Frascati (Roma) Italy\\
$^{6}$Faculty of Mathematics and Physics, University of Ljubljana, Jadranska 19, 1000 Ljubljana, Slovenia\\
$^{7}$Centre of Excellence Space-si, A\v sker\v ceva cesta 12, 1000 Ljubljana, Slovenia\\
$^{8}$Purple Mountain Observatory, Chinese Academy of Sciences, Nanjing 210008, China \\
$^{9}$Dark Cosmology Centre, Niels Bohr Institute, University of Copenhagen, Juliane Maries Vej 30, 2100, Copenhagen, Denmark \\
$^{10}$APC, Univ. Paris Diderot, CNRS/IN2P3, CEA/Irfu, Obs. de Paris, Sorbonne Paris Cit\'e, France\
}

\defcitealias{Sch10}{SCH10}

\begin{document}

\date{}

\pagerange{\pageref{firstpage}--\pageref{lastpage}} \pubyear{2011}

\maketitle

\label{firstpage}

\begin{abstract}
In this paper we compute rest-frame extinctions for the afterglows of a sample of \textit{Swift} $\gamma$-ray bursts complete in redshift. The selection criteria of the sample are based on observational high-energy parameters of the prompt emission and therefore our sample should not be biased against dusty sight-lines. It is therefore expected that our inferences hold for the general population of $\gamma$-ray bursts.
Our main result is that the optical/near-infrared extinction of $\gamma$-ray burst afterglows in our sample does not follow a single distribution. 87\% of the events are absorbed by less than 2\,mag, and 50\% suffer from less than 0.3-0.4\,mag extinction. The remaining 13\% of the afterglows are highly absorbed. The true percentage of $\gamma$-ray burst afterglows showing high absorption could be even higher since a fair fraction of the events without reliable redshift measurement are probably part of this class. These events may be due to highly dusty molecular clouds/star forming regions associated with the $\gamma$-ray burst progenitor or along the afterglow line of sight, and/or to massive dusty host galaxies.

No clear evolution in the dust extinction properties is evident within the redshift range of our sample, although the largest extinctions are at $z\sim1.5-2$, close to the expected peak of the star formation rate. Those events classified as dark are characterized, on average, by a higher extinction than typical events in the sample. A correlation between optical/near-infrared extinction and hydrogen-equivalent column density based on X-ray studies is shown although the observed $N_{\rm H}$ appears to be well in excess compared to those observed in the Local Group. Dust extinction does not seem to correlate with GRB energetics or luminosity.
\end{abstract}


\section{Introduction}

The study of the environments of long-duration $\gamma$-ray bursts (GRBs) is a subject of growing importance for its implications on many different research areas from GRB physics to host galaxy chemical evolution. The  cosmological nature of these events makes them  even more interesting, allowing to face the fundamental problem of determining how circumburst environmental parameters vary with the age of the universe \citep[see][for a review on this field]{Geh09}.

One possible way to obtain information about GRB environments along the line-of-sight (LOS) is to study their optical/near infrared (NIR) spectral energy distribution (SED). This allows researchers to derive information about the intrinsic spectrum in this spectral range and to study the rest-frame extinction curve, a precious source of information about dust formation (and possibly destruction) in high-redshift environments \citep[e.g.][]{Wax00,Fru01,Per02,Dra02,Per03,Str04,Che06b,Lu11}.

One of the first attempts to derive statistical information was carried out by \citet{Sch07} using \textit{Swift} data and their main results are still in agreement with our present knowledge: in most cases an extinction curve typical of the Small Magellanic Cloud (SMC) environment provides a good fit to the data and only in a few cases the extinction bump at about 2175\,\AA, typical of the Milky Way (MW) and the Large Magellanic Cloud (LMC) environments \citep[e.g.][]{Pei92} was singled out. \citet{Sch07} gave also support to the idea that afterglows undetected by \textit{Swift}-UVOT\footnote{http://www.swift.psu.edu/uvot}, in several cases classifiable as ``dark" bursts \citep[see e.g. ][and references therein for a discussion]{Mel11}, were those with high visual extinction \citep{Laz02}. \citet{Sch10}, with a larger sample of events, studied also the effect of soft X-ray absorption, which is usually quantified with the hydrogen-equivalent column density $N_{\rm H}$. They found that the ratio $N_{\rm H}/A_V$, with a large scatter, is significantly larger than what measured in the Magellanic Clouds or the MW. 

These studies however suffered from the limited spectral response and sensitivity of  \textit{Swift}-UVOT and the high level of inhomogeneity of the available observational data for GRB afterglows \citep{Kan10,Kan11}. The situation improved in recent years thanks to the growing attitude of teams involved in GRB follow-up to share their data for a better coverage of a specific event \citep[e.g.][]{Cov08,Jin12}, and in particular for the advent of advanced instruments such as GROND\footnote{http://www.mpe.mpg.de/$\sim$jcg/GROND/} equipping the 2.2\,m MPI/ESO telescope at La Silla (Chile). With its simultaneous 7-band imager, from the optical to the NIR, GROND allows to obtain homogeneous and uniformly calibrated data allowing to deal with afterglow SED determination with unprecedented reliability \citep[e.g.][]{Gre11}.

The search for highly extinguished GRB afterglows requires high-quality data of adequate spectral range, and indeed studies devoted to this subject \citep[e.g.][]{Per09,Kru11} could identify several events with moderate ($A_V \sim 1$\,mag) to high ($A_V > 2$\,mag) rest-frame extinction. \citet{Gre11}, further addressing the problem of the nature of dark GRBs, confirmed that for a large fraction of them (about 3/4) extinction is responsible for the faint optical fluxes, although still about up to 1/4 of the studied events are consistent with being high-redshift GRBs, with the optical flux depressed by intergalactic neutral hydrogen absorption.  

Even better results could in principle be obtained through optical/NIR spectroscopy, although the modest covered spectral range might sometimes negatively affect SED studies. \citet{Zaf11} studied a sample of 41 GRBs mainly observed with FORS\footnote{http://www.eso.org/sci/facilities/paranal/instruments/fors/} equipping the VLT. After modeling the SEDs from optical to X-rays by means of broken power-laws, they remarkably found that in about half of the studied events the SEDs require a break between the two bands consistent with a synchrotron origin, as required by the so-called GRB standard model \citep{Pir04}. In the remaining cases (apart from one outlier) a single power-law provides a satisfactory fit. In 63\% of the cases a SMC extinction curve is preferred, and the 2175\,\AA\ extinction bump is present just in 7\% of their sample. About a quarter of events is finally consistent with no absorption.

Any sample so far considered is in any case likely biased toward optically bright events. Therefore it is not probably totally surprising to see such a high fraction of virtually unreddened afterglows. This has clear impact on any inference about the nature of long GRB progenitors, usually supposed to be massive stars still in their star forming regions \citep{Scu11}. A likely improvement of SED modeling by means of spectral data will be possible when a large sample of afterglow observations, carried out with instruments such as the ESO X-shooter\footnote{http://www.eso.org/sci/facilities/paranal/instruments/xshooter/} \citep{Ver11}, capable to cover with low-to-medium resolution the whole optical/NIR range in one shot, will be available \citep[see, e.g.,][]{Del10,Wie11}.

From a different point of view, a further step toward the characterization of extinction properties of GRB LOSs should rely on statistical considerations based on a complete sample of events according to some selection criterion. To this aim we are studying in this paper the SEDs and extinction properties of a sample of 58 GRB afterglows (the BAT6 sample) described in \citet{Sal11}. All 58 GRBs in our sample have been selected to have the 1-s peak photon flux $P \ge 2.6$\,ph\,s$^{-1}$\,cm$^{-2}$ \citep[see][for further details]{Sal11} as detected by \textit{Swift}-BAT. The resulting sample has a redshift completeness level of $\sim 95$\% ($\sim 97$\% of the bursts have a constrained redshift). Several papers discussing various features of interest for GRB astrophysics for this sample have been delivered so far: X-ray spectra and absorbing columns are studied in \citet{Cam11}, the percentage of dark burst population is derived in \citet{Mel11}, spectral-energy correlations are discussed in \citet{Nav11} and \citet{Ghi12}, and general prompt/afterglow brightness correlations are derived in \citet{Dav11}.

The selection criterion based on high-energy brightness of the prompt emission avoids the introduction of biases related to the optical afterglow detection and, often, for later host galaxy identification. Yet, in general, a bright prompt emission implies a bright afterglow at any wavelength, although with a large scatter \citep{Dav11}. Therefore in a large fraction of cases the optical data available are adequate for the analysis. The very high level of redshift completeness makes it possible for the first time to obtain representative statistics of the extinction properties of GRB LOSs from nearby ($z \sim 0.1$) to far cosmological events ($z \sim 5.5$).

In Sect.\,\ref{sec:dtmth} we discuss the methodologies applied to the analysis of the events in our sample. In Sect.\,\ref{sec:rls} we report about our main results and in Sect.\,\ref{sec:cls} general conclusions are drawn. Details about the analysis for each event are discussed in Appendix\,\ref{sec:grbs}.

\section{Data and Methodology}
\label{sec:dtmth}

In order to compute the SED for as many events as possible included in our sample, we collected all the available data including those published only in GCN\footnote{http://gcn.gsfc.nasa.gov/} short communications. In a few cases, we had access to still unpublished data or derived a new calibration (see Appendix\,\ref{sec:grbs}).

The SED in the optical/NIR range is modeled as a power-law reddened by rest-frame extinction, $f_\nu \propto \nu^{-\beta} e^{-\tau(\nu (1+z))}$, where $\nu$ is the observed frequency, $z$ is the source redshift, $\tau(\nu)$ is the computed optical depth and $\beta$ is the spectral index. We applied three possible extinction curves, namely those typical for the MW, LMC and SMC, as modeled by \citet{Pei92}. MW extinction in our Galaxy is also considered \citep{Sch98}. We did not try a more complex modeling of the extinction curve \citep[see, e.g.][]{Cal94,Mai04,Gal10,LiLi10} since the inhomogeneity of the available data could easily introduce spurious results not justifying the increase in complexity in our statistical analysis. In addition, a reliable analysis of a grey extinction curves would be possible only for those events with optical/NIR and soft X-rays  lying on the same spectral segment, due to the degeneracy between optical/NIR spectral normalization and the break frequency location between these two spectral regions. This would have limited the sample size of our analysis to just a small fraction of the available events.
Temporal decays are modeled as power-laws, $f(t) \propto (t-t_0)^{-\alpha}$, where $t-t_0$ is the time after the burst and $\alpha$ the temporal index. In order not to artificially neglect the possible covariance between the spectral index and temporal decay when data are poorly sampled, temporal and spectral data are simultaneously fit. $\chi^2$ minimization is performed by using the downhill (Nelder-Mead) simplex algorithm as coded in the {\tt python}\footnote{http://www.python.org} {\tt scipy.optimize}\footnote{http://www.scipy.org/SciPyPackages/Optimize} library, v.\,0.10.0. Error search is carried out following \citet{Cas76} with $\beta$ and $E_{B-V}$ as parameters of interest. Throughout this paper the reported uncertainties are at $1\sigma$ level with the exception of hydrogen-equivalent column densities and X-ray spectral slopes, which are shown with 90\% errors. Upper limits are at 95\% confidence level. Lower limits have been derived by making use of the usually poor information available in the literature, as for instance rough upper limits published in GCNs. So, for lower limits, it is not possible to give a confidence level. They are anyway derived assuming the less demanding, in terms of required optical extinction, extrapolation of X-ray spectra to the optical band, i.e. introducing a spectral break just below the \textit{Swift}-XRT range. 

Our goal is to derive a reliable SED and extinction evaluation, therefore we did not try to model the whole light-curve of the analyzed events. We instead considered only the time interval required to firmly constrain the temporal index around the epoch when the most reliable photometric information is available. In several cases, not optimal calibration and/or inhomogeneity in the available data required to add a systematic calibration error in quadrature. We considered acceptable a fit if its null probability is better than 10\%. The spectral slope was derived both considering only optical/NIR data and adding a statical prior from the analysis of roughly simultaneous X-ray data. This is an alternative procedure compared to the simultaneous fit of X-ray and optical data, since in this latter case the solution is usually dominated by the much more numerous X-ray points, making fit results less sensitive to the information provided by the optical/NIR data. We imposed that $\beta_{\rm o} = \beta_{\rm X} - 0.5$ or $\beta_{\rm o} = \beta_{\rm X}$ including the 90\% error on the X-ray spectral slope. This is essentially equivalent to assume that the afterglow emission is due to synchrotron radiation and the cooling frequency is either outside or in between the X-ray and optical/NIR ranges \citep[see, e.g.,][and references therein]{Sar98,Zha04}. However, this choice does not imply that the afterglow evolution can be modeled in the framework of the cosmological fireball model \citep[see also][]{UhZa13}. Results for all extinction recipes are reported in Table\,\ref{tab:allres} (available only in electronic format) and details about each event fit and data selection are reported in Appendix \ref{sec:grbs}. In case more solutions are statistically acceptable we always chose the one obtained assuming a SMC extinction curve, for an easier comparison with most of the previous studies of GRB afterglow optical/NIR SEDs. Multiple acceptable solutions are typically possible only in case of low total extinction, when results based on different extinction recipes tend to be comparable.

For the analysis of the events in our sample we adopted the redshifts reported in \citet{Sal11}. X-ray spectral indices are taken from \citet{Mel11} and hydrogen-equivalent column densities from \citet{Cam11}. Energetics and luminosities for the prompt emission are taken from \citet{Nav11}.

The best fit results, which are the base for our further analysis, are reported in Table\,\ref{tab:bestres}, together with redshifts, hydrogen-equivalent column densities and spectral slopes.

\section{Results and discussion}
\label{sec:rls}

\begin{figure}
\begin{center}
\includegraphics[width=\columnwidth]{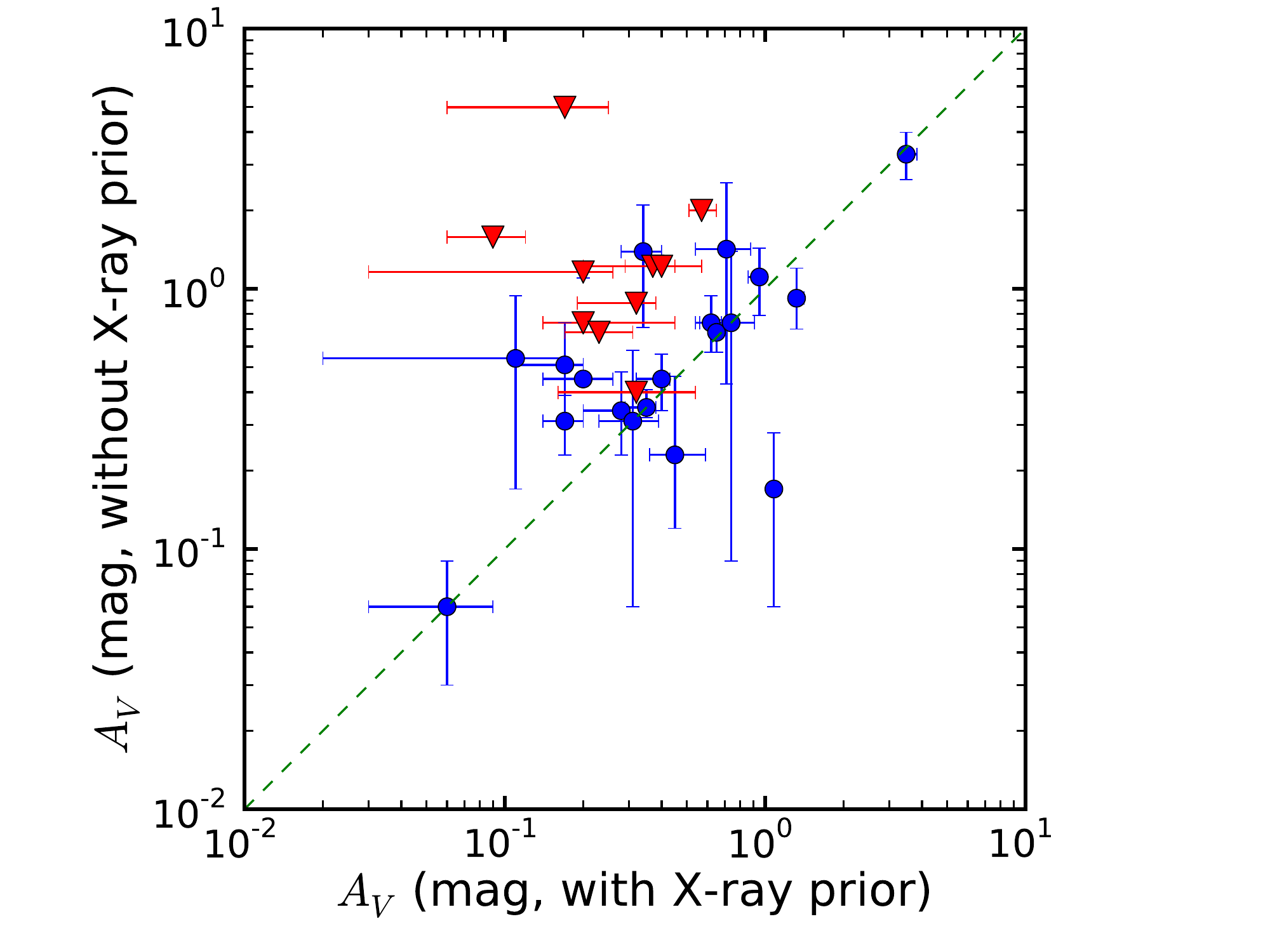}
\caption{Comparison between $A_V$ derived with and without assuming an X-ray prior. Only data for which at least one of the plotted quantities is not a limit are shown. Here and for later figures, blue (circle) symbols are actual measurements and red (triangles) symbols are upper or lower limits.}
\label{fig:avxav}
\end{center}
\end{figure}

In Table\,\ref{tab:bestres} we report results obtained leaving the spectral slope of the optical/NIR data completely free or constraining this parameter to be in agreement with the results obtained by the analysis of simultaneous X-ray data. However, only in a minority of cases optical data alone allowed us to obtain useful estimates of the rest-frame extinction. In those cases results from both analyses are in reasonable agreement (Fig.\,\ref{fig:avxav}). In the rest of the paper we will use only data obtained using the X-ray information. 

Our sample consists of 58 events, covering the redshift range from $\sim 0.1$ to $\sim 5.5$. Rest-frame extinctions or limits on it could be estimated for 53 events ($\sim 91 $\%).
For the remaining cases no information could be obtained due to the poor quality of optical data, X-ray data, or the lack of a redshift. In at least 8 cases ($\sim 15$\%) an extinction curve presenting a bump at $\sim 2175$\,\AA, i.e. LMC or MW, is required. \citet{Zaf11}, through GRB afterglow spectroscopy, measured this percentage to be about $7$\%, although in their sample high extinction events would likely be under-represented for observational biases. Broad-band photometry is clearly not the best tool to single out the extinction bump, in particular for low total extinction, and the true percentage is likely to be affected by the application of possible different extinction recipes \citep[e.g.][]{Cal94,Mai04,LiLi10}. In a large fraction of cases, a chromatic extinction law is definitely required by the data. 

\begin{figure*}
\begin{center}
\includegraphics[width=\textwidth]{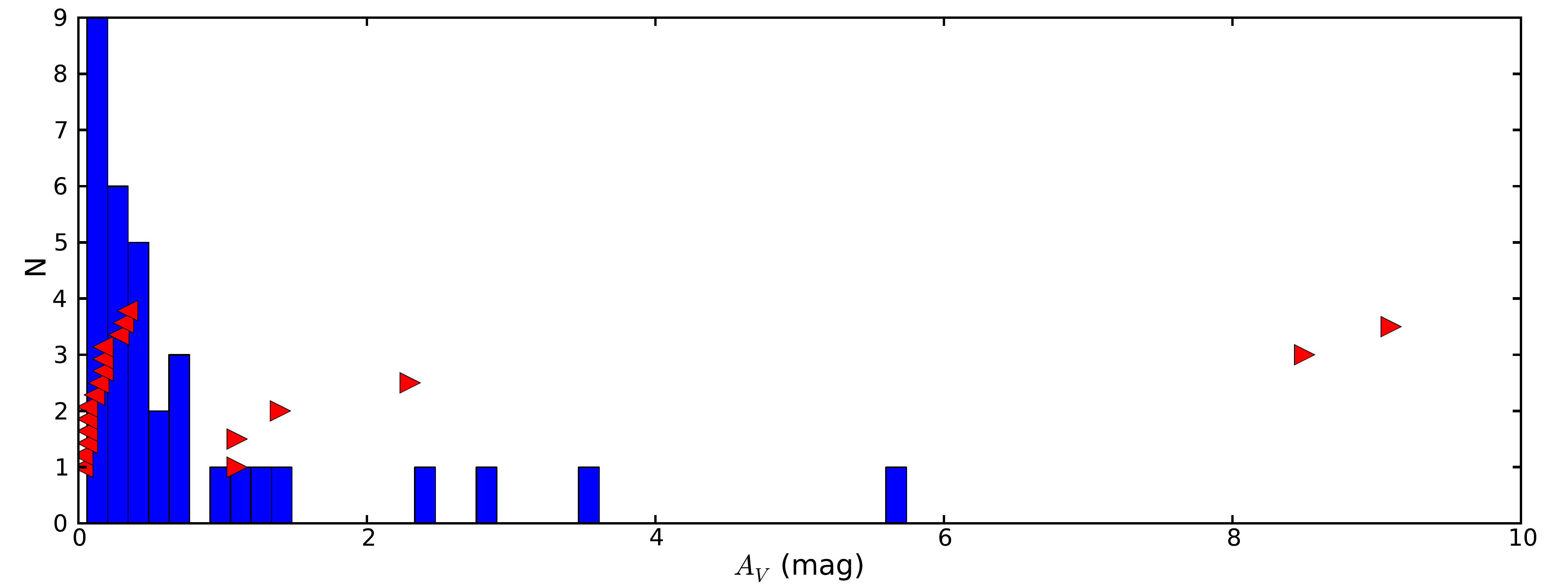}
\caption{Histogram of the computed rest-frame $A_V$. Upper and lower limits are also reported. 87\% of the GRB afterglows suffer from an extinction lower than about 2\,mag, and 50\% lower than 0.3-0.4\,mag. About 13\% of the population is characterized by a large extinction.}
\label{fig:avhist}
\end{center}
\end{figure*}

The distribution of rest-frame extinctions is shown in Fig.\,\ref{fig:avhist} and resembles analogous histograms discussed by \citet{Kan10} and \citet{Gre11}. However, since our sample is selected with purely observational criteria of the prompt $\gamma$-ray emission, and it is highly complete in redshift, is likely to provide an unbiased view of the true ``parent" distribution of rest-frame extinction for cosmological GRBs. The distribution seems to be bimodal, with a smooth distribution clustered at virtually zero extinction (including the upper limits) and several events (including the lower limits) at higher, roughly $A_V \ge 2$\,mag, extinction. 87\% of the GRB afterglows (46/53) have an extinction smaller than about 2\,mag, and 50\% smaller than 0.3-0.4\,mag. The distribution is indeed strongly peaked at low extinction. The remaining 13\% of events are on the contrary suffering from high extinction. They appear to follow, within the limits of the size of the sample, a different distribution being essentially inconsistent with a simple extrapolation of low-extinction events at more than 3$\sigma$ level, if we model the low-extinction peak with a Gaussian or a Poissonian function. The percentage of events with large extinction could be even higher since the few events with no redshift measurements in our sample are likely part of this class. The true percentage of GRB afterglows showing high extinction is therefore probably between 15-20\%, well in the range of the estimated percentage of dark bursts derived by means of different criteria and in different samples\citep{deP03,Jak04,Rol05,van09,Gre11,Mel11}.

The shape of the low-extinction event distribution is roughly consistent with what could be expected in typical Galactic molecular clouds, although the amount of extinction is substantially lower than expected \citep[e.g.][]{RePr02,Ver04}. The 15-20\% of events with high extinction could be part of a different population of GRBs born in very dusty molecular clouds \citep{Pro09,Per09} and/or hosted by dusty, more massive, galaxies than the average of the host GRB population \citep[e.g.][]{Kru11,Ros12,Per13}.


\begin{figure}
\begin{center}
\includegraphics[width=\columnwidth]{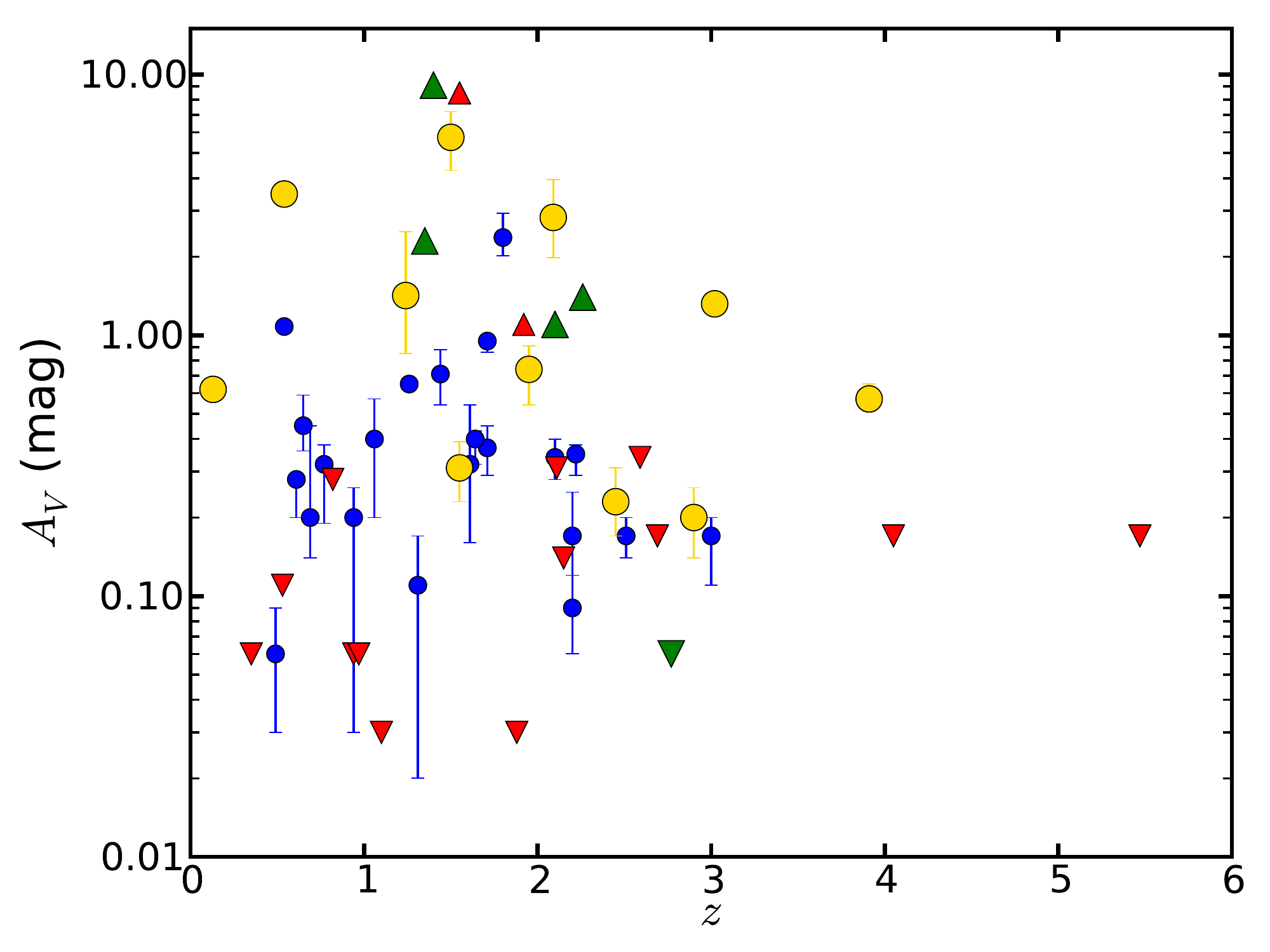}
\caption{Extinction versus redshift. No clear relation appears to be present in the redshift range covered by our sample, although the largest extinctions are at $z\sim1.5-2$, close to the expected peak of the star formation rate \citep[$z \sim 2 - 3$, ][]{HoBe06}. Bigger symbols (gold: measurements, green: limits) show data for dark bursts in our sample according to \citet{Mel11}. }
\label{fig:avz}
\end{center}
\end{figure}

The derived rest-frame extinctions do not show a clear redshift dependence at least up to $z \sim 4$, where we have a sufficient number of events (Fig.\,\ref{fig:avz}). The most stringent upper limits are at low redshift, but this is probably an observational bias since the best optical data are found for those events. The largest extinctions are however at  $z\sim1.5-2$, close to the expected peak of the star formation rate \citep[$z \sim 2 - 3$, ][]{HoBe06}. This might have interesting consequences although the observed extinction in GRB afterglows is related to their LOSs and their physical properties do not necessarily correlate with global properties of their host galaxies. In the same plot we show the position in the $A_V$ vs $z$ plane of the dark bursts in our sample \citep{Mel11}. It is clear that, as expected, most dark bursts are characterized by a substantial optical extinction, in agreement with previous studies \citep[e.g.][]{Gre11,Kru11}. An interesting exception is provided by GRB\,081222 that is classified as dark \citep{Mel11}. Yet, it seems to be characterized by a very low dust extinction although the classification of this event as dark is questionable, as discussed in \citet{Mel11}. On the contrary, GRB\,060306 is characterized by a high extinction and it was not classified as dark. For this event only upper limits in the NIR are available (see Appendix\,\ref{sec:grbs}), and therefore even its classification is very uncertain. In any case, the classification of an event as dark always requires a careful comparison between optical and X-ray emission, and therefore the amount of optical extinction is only one of the important factors involved in the classification.

\begin{figure}
\begin{center}
\includegraphics[width=\columnwidth]{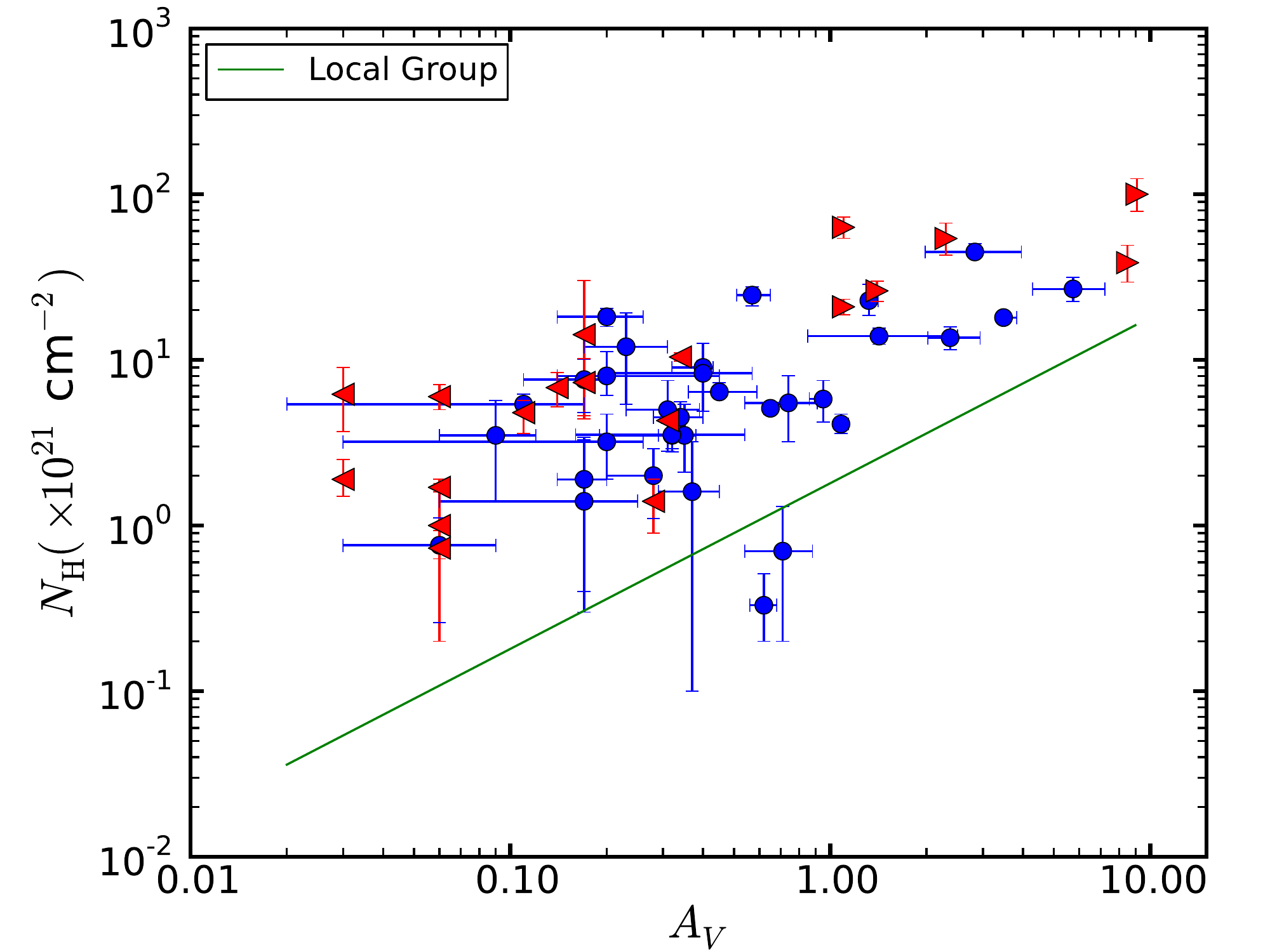}
\caption{Relation between $A_V$ and $N_{\rm H}$ derived by X-ray analysis. Typical dust-to-gas ratio for LG environments is also shown. Only data for which at least one of the plotted quantities is not a limit have been shown. No correction for metallicity different from Solar was applied to the observed $N_{\rm H}$.}
\label{fig:avnh}
\end{center}
\end{figure}

The relation between $A_V$ and $N_{\rm H}$ derived by X-ray analysis has been widely discussed in the literature \citep[e.g.][]{Str04,Wat07,Sch10,Zaf11,Wat13}. Taking into account the metallicity effect on the observed hydrogen-equivalent column densities there is a general consensus about a low dust-to-gas ratio ($\sim 10$\%) compared to typical Local Group (LG) values. Our sample (Fig.\,\ref{fig:avnh}) indeed singles out a well defined trend with the highest rest-frame extinctions derived for the highest hydrogen-equivalent absorptions, roughly following the relation $N_{\rm H}/A_V \sim 1.6 \times 10^{22}$\,cm$^{-2}$\,mag$^{-1}$. The typical LG dust-to gas ratio \citep{Wel12} for Solar metallicity is also shown. Nonetheless, up to $A_V \sim 1$\,mag the trend is not evident and the there is a scattered distribution. The hint for two populations found from the $A_V$ distribution (Fig.\,\ref{fig:avhist}) might be found also in the $N_{\rm H}/A_V$ plan. The most absorbed afterglows (with $A_V \ge 1$\,mag) are all characterized by the highest $N_{\rm H}$ values, i.e. larger than $10^{22}$\,cm$^{-2}$.

\begin{figure}
\begin{center}
\includegraphics[width=\columnwidth]{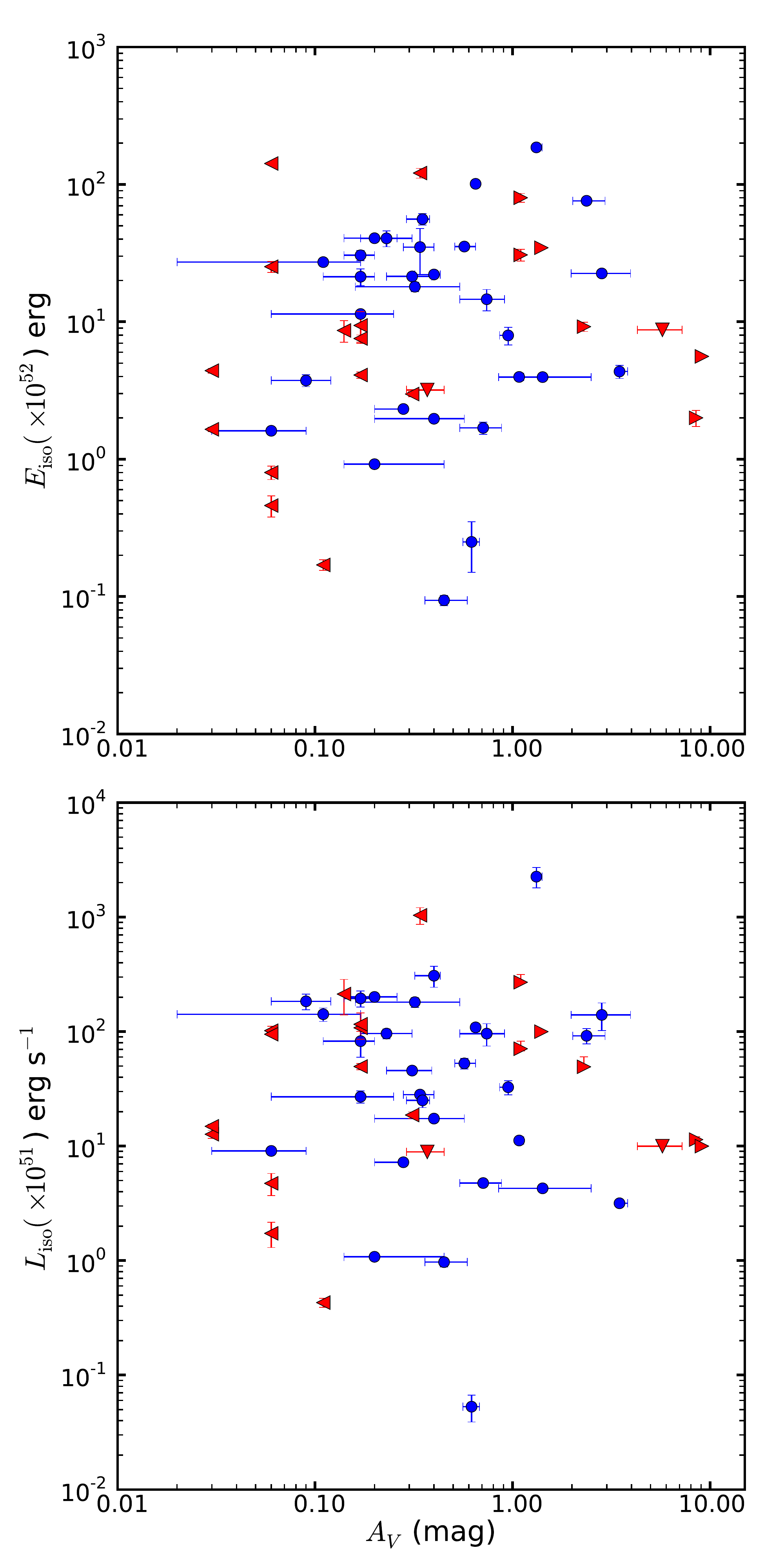}
\caption{Extinction versus $E_{\rm iso}$ (upper panel) and $L_{\rm iso}$ (lower panel). The amount of extinction does not seem to be affected by the energetics or luminosity of the GRB.}
\label{fig:aveisoliso}
\end{center}
\end{figure}

Hydrogen-equivalent column densities derived through analysis of GRB afterglow X-ray data are actually due to photoelectric absorption by inner shells of elements as O, Si, S, Fe and He \citep[see][and references therein]{Wat13}. The relation with hydrogen is obtained typically assuming a gas composition in Solar proportion and Solar metallicity. The comparison with dust absorption depends also on the dust-to-metal ratio \citep[e.g.][]{Sav03,ZaWa13}. None of these parameters can, in general, be robustly determined by observations or predicted theoretically for the whole sample, making any firm inference about the relation between the X-ray $N_{\rm H}$ and $A_V$ still difficult. Dust sublimation by the GRB emission \citep[e.g.][]{Wax00} is usually invoked as a possible explanation for the low dust content along GRB LOSs, while the effect on the measured $N_{\rm H}$ can be less important since (mildly) ionized metals can still produce absorption in low-resolution soft X-ray spectra \citep[e.g.][]{Per03}. However, as shown in Fig.\,\ref{fig:aveisoliso}, the energetics or luminosity of the GRB is not correlated with the amount of dust absorption, as it could be expected in case of sizable dust sublimation.

\begin{figure}
\begin{center}
\includegraphics[width=\columnwidth]{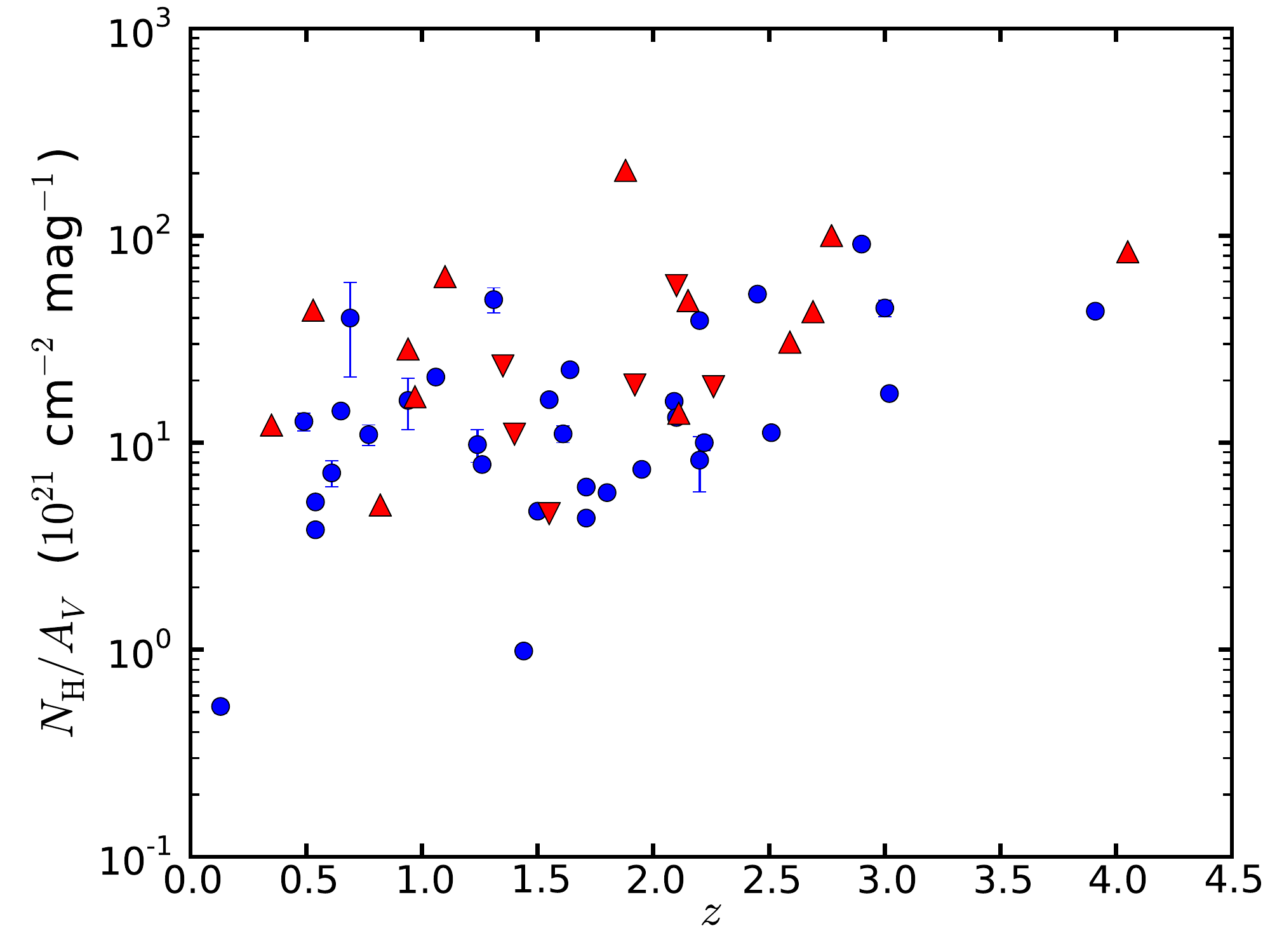}
\caption{$N_{\rm H}/A_V$ ratio versus redshift. A lack of low hydrogen-equivalent column densities at high redshifts mimics an increase of $N_{\rm H}/A_V$ \citep[see][]{Cam10,Cam11}.}
\label{fig:avznh}
\end{center}
\end{figure}

\citet{Wat13} reported an evolution in redshift for the $N_{\rm H}/A_V$ relation, with higher values at high redsfhits. Based on our sample (Fig.\,\ref{fig:avznh}) we can say that this is a direct consequence of lack of low $N_{\rm H}$ values at high redshift \citep[see][]{Cam10, Cam11}, whereas $A_V$ does not evolve with redshift (see Fig.\,\ref{fig:avz}). The physical reason of the preference for high $N_{\rm H}$ values at high redshifts and its significance have still to be determined.

In \citet{Cam11} this phenomenon is attributed to the increasing absorption by intervening systems in higher redshift GRBs and therefore it would not be related to the progenitor environment (see, however, \citealt{Wat13} and \citealt{Sta13}). The effect is important. The $N_{\rm H}$ contribution due to intervening systems, derived by a rough estimate from, e.g. Fig.\,2 in \citet{Cam10}, at $z \sim  3$ is  $\sim 6 \times 10^{20}$\,cm$^{-2}$ and at $z \sim 6$ rises to $\sim 2 \times 10^{22}$\,cm$^{-2}$. The subtraction of this contribution from the measured hydrogen-equivalent column densities does not however make the relation with the optical extinction more defined. The intrinsic uncertainties in deriving this relation added to the uncertainties associated to the derived $N_{\rm H}$ make it actually difficult to derive firmer conclusions.

\begin{figure}
\begin{center}
\includegraphics[width=\columnwidth]{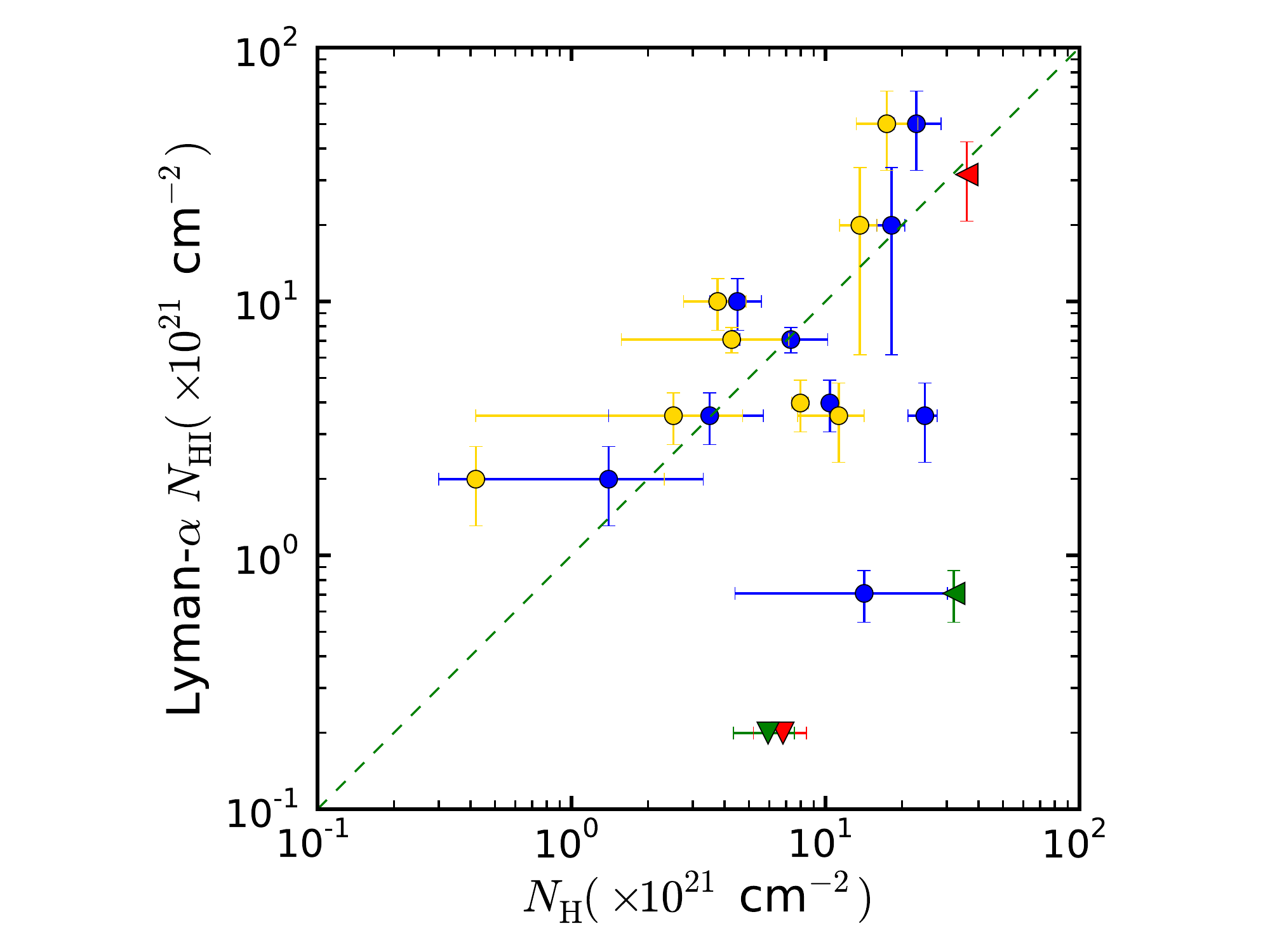}
\caption{Relation between $N_{\rm H}$ from X-ray spectral analysis and $N_{\rm HI}$ derived by measuring Lyman-$\alpha$. Blue and red symbols are as in Fig.\,\ref{fig:avxav}. Gold and green symbols are measurements and upper limits after subtracting from the X-ray hydrogen-equivalent column density a contribution from intervening systems \citep[see][]{Cam11}.}
\label{fig:avnhlnh}
\end{center}
\end{figure}

The hydrogen column density can also be measured more directly when Lyman-$\alpha$ is visible in GRB afterglow spectra. This practically requires that events be at a redshift sufficiently high to have the Lyman-$\alpha$ transition in the useful spectral range of ground-based instruments, i.e. from the ultraviolet atmospheric cutoff at $\sim 3500$\,\AA\  upwards. The redshift limit therefore is approximately $z > 2$. This limits the number of events for which Lyman-$\alpha$ has been measured. In our sample there are just 11 such events. Data are from \citet[][and references therein]{Fyn09} but for GRB\,081203A \citep{Kui09}. In Fig.\,\ref{fig:avnhlnh} we show the relation between $N_{\rm H}$ based on Lyman-$\alpha$ observation ($N_{\rm HI}$) and $N_{\rm H}$ based on X-ray analysis. 
If both $N_{\rm HI}$ and the gas responsible for the X-ray absorption come from the diffuse interstellar medium of the host galaxy, the two values should be correlated. For some of the events in our sample, rather surprisingly, the agreement is satisfactory. Possibly, for the sub-sample of events with both measurements, metallicity effects on the observed X-ray based $N_{\rm H}$ are not dominant.  Nonetheless, as already discussed, the X-ray values are calculated using Solar metallicity, which might not correspond to the host metallicity, and past studies using larger samples found an excess of the X-ray hydrogen-equivalent column density compared to $N_{\rm HI}$. It is interesting to note that if we again subtract to the X-ray values the mean contribution in the observed column density resulting from intervening systems as a function of redshift as empirically calculated by \citet{Cam11}, the relation between $N_{\rm H}$ and $N_{\rm HI}$ does become tighter (see Fig.\,\ref{fig:avnhlnh}), as expected if the gas responsible for both absorptions comes only from the GRB host. This can be a further hint, within the limits of the small available sample, for a possible contribution of the intervening systems to the hydrogen-equivalent column density determined from the X-ray.

%

\begin{table*}
\centering
\begin{minipage}{180mm}
\caption{Spectral and absorption parameters for GRBs in our sample. For each event we list the redshift, the X-ray spectral index, the X-ray hydrogen-equivalent absorbing column, the optical/NIR spectral index and the amount of rest-frame reddening in the visual band.}
\label{tab:bestres}
\begin{tabular}{lccccclccl}
\hline
Event                  & $z$  & $\beta_{\rm X}$                        & $N_{\rm H}$        &       \multicolumn{3}{c}{No X-ray prior} &  \multicolumn{3}{c}{With X-ray prior} \\
			&	  &					  & $10^{21}$\,cm$^{-2}$	&   $\beta_{\rm o}$   & $A_V$                 & Ext. curve & $\beta_{\rm o}$   & $A_V$                 & Ext. curve \\
\hline
GRB\,050318    & 1.44 & $0.95^{+0.07}_{-0.06}$ & $0.7^{+0.6}_{-0.5}$   & $-2.00_{-4.90}^{+3.89}$ & $1.42_{-0.99}^{+1.13}$ & SMC & $0.89_{-0.00}^{+0.13}$    & $0.71^{+0.17}_{-0.17}$ & SMC \\
GRB\,050401    & 2.90 & $0.83^{+0.15}_{-0.14}$ & $18.2^{+2.3}_{-2.3}$ &  $-0.42_{-0.73}^{+0.65}$ & $0.45_{-0.25}^{+0.28}$ & SMC & $0.19^{+0.29}_{-0.00}$    & $0.20^{+0.06}_{-0.11}$ & SMC \\
GRB\,050416A & 0.65 & $1.11^{+0.11}_{-0.14}$ & $6.4^{+0.9}_{-0.5}$   & $1.04^{+0.25}_{-0.33}$ & $0.23_{-0.11}^{+0.23}$ & SMC & $0.72_{-0.17}^{+0.00}$    & $0.45^{+0.14}_{-0.09}$ & SMC \\ 
GRB\,050525A & 0.61 & $1.08^{+0.15}_{-0.13}$ & $2.0^{+0.9}_{-0.9}$   & $0.27^{+0.48}_{-0.52}$ & $0.34^{+0.14}_{-0.11}$ & SMC & $0.45_{-0.00}^{+0.28}$    & $0.28^{+0.00}_{-0.08}$ & SMC \\
GRB\,050802   & 1.71 &  $0.89^{+0.04}_{-0.07}$ & $1.6^{+1.6}_{-1.5}$    & $0.34_{-1.42}^{+1.10}$ & $< 1.22$  & SMC & $0.34_{-0.02}^{+0.09}$ & $0.37_{-0.08}^{+0.08}$   & SMC \\
GRB\,050922C & 2.20 & $1.25^{+0.06}_{-0.07}$  & $3.5^{+2.2}_{-2.1}$  & $-0.30^{+1.47}_{-4.74}$ & $< 1.58$ & SMC & $0.68^{+0.13}_{-0.00}$ & $0.09_{-0.03}^{+0.03}$ & SMC \\
GRB\,060206 & 4.05 & $1.30^{+0.57}_{-0.53}$ & $14.2^{+16.0}_{-9.8}$ & $0.46^{+0.41}_{-0.39}$ & $< 0.25$ & SMC &   $0.46^{+0.41}_{-0.19}$ & $< 0.17$  & SMC \\
GRB\,060210 & 3.91 & $1.08^{+0.05}_{-0.05}$ & $24.6^{+2.9}_{-3.5}$   & $4.47^{+0.28}_{-9.51}$ & $< 2.00$ & SMC & $1.13^{+0.00}_{-0.10}$ & $0.57_{-0.06}^{+0.08}$ & SMC \\
GRB\,060306 & 1.55\footnote{This redshift was revised by \citet{Jak12} and \citet{Per13}.} &  $1.62^{+0.28}_{-0.26}$ & $38.6^{+10.5}_{-9.1}$  & & & & 0.86 & $> 8.5$ & SMC  \\
GRB\,060614 & 0.13 & $0.89^{+0.06}_{-0.04}$ & $0.33^{+0.18}_{-0.13}$ & $0.18^{+0.20}_{-0.23}$ & $0.74_{-0.17}^{+0.20}$ & SMC & $0.35^{+0.09}_{-0.00}$ & $0.62_{-0.06}^{+0.06}$ & SMC\\
GRB\,060814 & 1.92 & $1.13^{+0.07}_{-0.07}$ & $20.9^{+2.3}_{-2.2}$ & & & & 0.56 & $> 1.1$ & SMC \\
GRB\,060904A &  & $0.28^{+0.74}_{-0.47}$ & $> 2.07$ \\
GRB\,060908 & 1.88 & $1.42^{+0.32}_{-0.36}$ & $6.2^{+2.8}_{-2.5}$  & $0.22^{+0.22}_{-0.26}$ & $< 0.20$ & SMC & $0.56^{+0.02}_{-0.00}$ & $< 0.03$ & SMC\\
GRB\,060912A & 0.94 & $0.71^{+0.19}_{-0.25}$ & $3.2^{+1.5}_{-1.3}$ & $-0.08^{+1.23}_{-1.42}$ & $< 1.16$ & SMC  & $0.46^{+0.44}_{-0.00}$ & $0.20_{-0.17}^{+0.06}$ & SMC\\
GRB\,060927 & 5.47 &  $0.96^{+0.33}_{-0.25}$ & $< 36$  & $0.75^{+0.41}_{-2.35}$ & $< 1.33$ & SMC & $0.75^{+0.41}_{-0.04}$ & $< 0.17$ & SMC \\
GRB\,061007 & 1.26 &  $1.01^{+0.09}_{-0.06}$ & $5.1^{+0.3}_{-0.3}$ & $0.87^{+0.27}_{-0.24}$ & $0.68_{-0.11}^{+0.08}$ & LMC &  $0.97^{+0.13}_{-0.02}$ & $0.65_{-0.03}^{+0.03}$ & LMC  \\ 
GRB\,061021 & 0.35 & $1.00^{+0.04}_{-0.05}$ & $0.73^{+0.2}_{-0.1}$ &  $0.04^{+0.42}_{-0.60}$ & $< 0.42$ & SMC &  $0.45^{+0.09}_{-0.00}$ & $< 0.06$ & SMC \\
GRB\,061121 & 1.31 & $0.91^{+0.06}_{-0.06}$ & $5.4^{+0.8}_{-0.5}$ & $-1.15^{+1.21}_{-1.41}$ & $0.54_{-0.37}^{+0.40}$ & SMC &  $0.35_{-0.00}^{+0.12}$ & $0.11_{-0.09}^{+0.06}$ & SMC \\
GRB\,061122A & 2.09 & $0.95^{+0.07}_{-0.06}$ & $44.8^{+5.4}_{-3.0}$ & & & & $0.45^{+0.07}_{-0.06}$ & $2.83^{+1.13}_{-0.85}$ & SMC \\
GRB\,070306 & 1.50 & $0.95^{+0.07}_{-0.06}$ & $26.8^{+4.7}_{-4.3}$  & & & & $0.39_{-0.00}^{+0.13}$ & $5.74_{-1.45}^{+1.48}$ & SMC \\
GRB\,070328 & $< 4$ & $0.95^{+0.08}_{-0.08}$\\
GRB\,070521 & 1.35 & $1.03^{+0.15}_{-0.13}$ & $54^{+13}_{-11}$ & & & & 0.40 & $> 2.3$ & SMC \\
GRB\,071020 & 2.15 & $0.89^{+0.16}_{-0.14}$ & $6.8^{+1.6}_{-1.6}$ & $0.58^{+0.28}_{-1.71}$ & $< 1.13$ & SMC & $0.75^{+0.20}_{-0.00}$ & $< 0.14$ & SMC \\
GRB\,071112C & 0.82 & $0.79^{+0.21}_{-0.27}$ & $1.4^{+0.5}_{-0.5}$ & $0.48^{+0.19}_{-0.37}$ & $< 0.28$ & SMC & $0.49^{+0.01}_{-0.37}$ & $< 0.28$ & SMC  \\
GRB\,071117 & 1.33 & $1.09^{+0.13}_{-0.19}$   & $10.9^{+2.1}_{-3.1}$ \\
GRB\,080319B & 0.94 & $0.82^{+0.06}_{-0.06}$   & $1.7^{+0.1}_{-0.1}$ & $0.22^{+0.15}_{-0.15}$ & $0.09_{-0.06}^{+0.06}$ & SMC &   $0.26^{+0.10}_{-0.00}$ & $< 0.06$ & SMC  \\
GRB\,080319C & 1.95 & $0.97^{+0.28}_{-0.23}$   & $5.5^{+2.5}_{-2.3}$ & $0.48^{+1.66}_{-1.64}$ & $0.74_{-0.65}^{+0.65}$ & SMC  & $0.48^{+0.27}_{-0.24}$ & $0.74_{-0.20}^{+0.17}$ & SMC \\
GRB\,080413B & 1.10 &  $0.97^{+0.05}_{-0.07}$   & $1.9^{+0.6}_{-0.4}$ & $0.13^{+0.08}_{-0.08}$ & $0.11_{-0.03}^{+0.03}$ & SMC & $0.40^{+0.01}_{-0.00}$ & $< 0.03$ & SMC \\
GRB\,080430 & 0.77 & $1.06^{+0.06}_{-0.07}$ & $3.5^{+0.8}_{-0.6}$ & $0.50^{+0.95}_{-1.27}$ & $< 0.88$ & LMC & $0.49^{+0.13}_{-0.00}$ & $0.32^{+0.06}_{-0.13}$ & LMC  \\
GRB\,080602 & 1.40 & $0.90^{+0.12}_{-0.13}$   & $6.7^{+2.4}_{-2.1}$ \\
GRB\,080603B & 2.69 & $0.87^{+0.26}_{-0.21}$  & $7.3^{+2.9}_{-2.7}$ & $0.22^{+0.55}_{-0.74}$ & $< 0.45$ & SMC & $0.19^{+0.44}_{-0.03}$ & $< 0.17$ & SMC \\
GRB\,080605 & 1.64 & $0.86^{+0.11}_{-0.16}$ & $9.0^{+0.9}_{-0.9}$ & $0.65^{+0.24}_{-0.22}$ & $0.45_{-0.11}^{+0.11}$ & SMC &  $0.74^{+0.20}_{-0.04}$ & $0.40_{-0.08}^{+0.03}$ & SMC \\
GRB\,080607 & 3.02 & $1.13^{+0.06}_{-0.11}$ &  $22.8^{+5.7}_{-4.2}$ & $1.73^{+0.27}_{-0.29}$ & $0.92^{+0.28}_{-0.22}$ & MW & $1.19^{+0.00}_{-0.11}$ & $1.32^{+0.09}_{-0.06}$ & MW \\ 
GRB\,080613B &  & $1.39^{+1.28}_{-0.87}$ &  $> 0.5$ \\
GRB\,080721 & 2.59 & $0.91^{+0.05}_{-0.05}$ &  $10.4^{+0.6}_{-0.6}$ & $-4.13^{+5.13}_{-7.30}$ & $< 2.97$ & SMC & $0.86^{+0.10}_{-0.00}$ & $< 0.34$ & SMC \\
GRB\,080804 & 2.20 & $0.97^{+0.12}_{-0.12}$ &  $1.4^{+1.9}_{-1.1}$ & $-8.45^{+10.10}_{-18.18}$ & $< 4.98$ & SMC & $0.37^{+0.22}_{-0.02}$ & $0.17^{+0.08}_{-0.11}$ & SMC \\ 
GRB\,080916A & 0.69 & $1.07^{+0.13}_{-0.18}$ & $8.0^{+3.2}_{-1.9}$ & $1.47^{+0.09}_{-0.43}$ & $< 0.74$ & SMC & $1.20^{+0.00}_{-0.31}$ & $0.20^{+0.25}_{-0.06}$ & SMC \\
GRB\,081007 & 0.53 & $1.04^{+0.10}_{-0.18}$ & $4.8^{+0.9}_{-1.2}$ & $0.43^{+0.33}_{-0.38}$ & $0.31^{+0.25}_{-0.23}$ & SMC & $0.86^{+0.07}_{-0.00}$ & $< 0.11$ & SMC \\
GRB\,081121 & 2.51 & $0.95^{+0.08}_{-0.08}$ & $1.9^{+1.6}_{-1.5}$ & $-0.31^{+0.43}_{-0.41}$ & $0.31^{+0.08}_{-0.08}$ & SMC & $0.37^{+0.12}_{-0.00}$ & $0.17^{+0.03}_{-0.03}$ & SMC \\
GRB\,081203A & 2.10 & $1.14^{+0.09}_{-0.10}$ & $4.5^{+1.1}_{-1.0}$  & $-5.92^{+4.18}_{-4.51}$ & $1.39^{+0.71}_{-0.68}$ & SMC & $0.67^{+0.12}_{-0.07}$ & $0.34^{+0.06}_{-0.06}$ & SMC \\
GRB\,081221 & 2.26 & $1.50^{+0.12}_{-0.11}$ & $26.1^{+3.8}_{-3.6}$  & & & & 0.89 & $> 1.4$ & SMC \\ 
GRB\,081222 & 2.77 & $1.03^{+0.07}_{-0.06}$ & $6.0^{+1.1}_{-1.0}$ & $-0.01^{+0.52}_{-0.54}$ & $< 0.54$ & SMC & $0.47^{+0.12}_{-0.00}$ & $< 0.06$ & SMC \\
GRB\,090102 & 1.55 & $0.78^{+0.06}_{-0.08}$ & $5.0^{+2.5}_{-2.2}$ & $0.27^{+0.54}_{-0.59}$ & $0.31^{+0.27}_{-0.25}$ & SMC & $0.27^{+0.09}_{-0.05}$ & $0.31^{+0.08}_{-0.08}$ & SMC \\
GRB\,090201 & $2.1$\footnote{This redshift is reported in \citet{Kru13b}.} & $0.97^{+0.11}_{-0.10}$ & $63.1^{+9.6}_{-8.9}$ & & & & 0.49 & $> 1.1$ & SMC \\
GRB\,090424 & 0.54 & $0.95^{+0.10}_{-0.09}$ & $4.1^{+0.6}_{-0.5}$ & $1.41^{+0.20}_{-0.15}$ & $0.17^{+0.11}_{-0.11}$ & SMC & $0.55^{+0.00}_{-0.06}$ & $1.08^{+0.06}_{-0.03}$ & MW \\
GRB\,090709A & 1.80\footnote{This redshift was revised by \citet{Per13}.} & $0.87^{+0.07}_{-0.07}$ & $13.6^{+2.2}_{-2.1}$ & $0.24^{+2.93}_{-7.48}$ & $< 15.8$ & LMC & $0.44^{+0.00}_{-0.14}$ & $2.37^{+0.57}_{-0.35}$ & LMC \\
GRB\,090715B & 3.0 & $1.04^{+0.09}_{-0.09}$ & $7.6^{+2.5}_{-2.8}$ & $-1.09^{+0.85}_{-0.91}$ & $0.51^{+0.23}_{-0.20}$ & SMC & $0.45^{+0.00}_{-0.18}$ & $0.17^{+0.03}_{-0.06}$ & SMC \\
GRB\,090812 & 2.45 & $0.95^{+0.07}_{-0.06}$ & $12.0^{+7.2}_{-6.6}$ & $0.58^{+0.85}_{-1.01}$ & $< 0.68$ & SMC &  $0.52^{+0.00}_{-0.13}$ & $0.23_{-0.06}^{+0.08}$ & SMC \\
GRB\,090926B & 1.24 & $0.95^{+0.07}_{-0.06}$ & $13.9^{+1.6}_{-1.5}$ & & & &  & $(1.42^{+1.08}_{-0.57})$\footnote{This figure is reported in \citet{Gre11}, although the related photometry has not been published.} & SMC  \\
GRB\,091018 & 0.97 & $1.18^{+0.18}_{-0.23}$ & $1.0^{+0.9}_{-0.8}$ & $0.57^{+0.01}_{-0.06}$ & $< 0.06$ & SMC & $0.57^{+0.01}_{-0.06}$ & $< 0.06$  & SMC \\ 
\hline
\end{tabular}
\end{minipage}
\end{table*}

\begin{table*}
\centering
\begin{minipage}{180mm}
\setcounter{footnote}{5}
\contcaption{}
\begin{tabular}{lccccclccl}
\hline
GRB\,091020 & 1.71 & $1.11^{+0.05}_{-0.06}$ & $5.8^{+1.7}_{-1.6}$ & $0.23^{+0.65}_{-0.64}$ & $1.11^{+0.32}_{-0.32}$ & SMC & $0.55^{+0.11}_{-0.00}$ & $0.95^{+0.03}_{-0.09}$ & SMC \\
GRB\,091127 & 0.49 &  $0.80^{+0.11}_{-0.11}$ & $0.76^{+0.35}_{-0.50}$  & $0.28^{+0.03}_{-0.02}$ & $0.06^{+0.03}_{-0.03}$ & SMC &  $0.29^{+0.00}_{-0.04}$ & $0.06^{+0.03}_{-0.03}$ & SMC \\
GRB\,091208B & 1.06 & $0.94^{+0.13}_{-0.08}$ & $8.3^{+4.3}_{-3.4}$ &  $0.94^{+0.83}_{-1.19}$ & $< 1.22$ & SMC & $0.94^{+0.13}_{-0.08}$ & $0.40_{-0.20}^{+0.17}$ & SMC \\
GRB\,100615A & 1.40\footnote[\value{footnote}]{This redshift is reported in \citet{Kru13}.}  & $1.20^{+0.30}_{-0.28}$ & $100^{+24.0}_{-21.0}$ & & & & 0.76 & $> 9.1$ & SMC \\
GRB\,100621A & 0.54 & $1.40^{+0.13}_{-0.12}$ & $18.0^{+1.2}_{-1.1}$ & $1.19^{+0.64}_{-0.65}$ & $3.29^{+0.70}_{-0.66}$ & LMC & $1.03^{+0.00}_{-0.25}$ & $3.48^{+0.35}_{-0.09}$ & LMC \\
GRB\,100728B & 2.11 & $1.08^{+0.17}_{-0.18}$ & $4.3^{+3.1}_{-2.5}$ & $0.70^{+0.31}_{-0.46}$ & $< 0.48$ & SMC & $0.70^{+0.05}_{-0.30}$ & $< 0.31$ & SMC \\
GRB\,110205A & 2.22 & $1.13^{+0.09}_{-0.09}$ & $3.5^{+1.9}_{-1.4}$  & $0.55_{-0.12}^{+0.10}$ & $0.35_{-0.03}^{+0.06}$ & LMC & $0.55_{-0.01}^{+0.14}$ & $0.35_{-0.06}^{+0.03}$ & LMC\\
GRB\,110503A & 1.61 & $0.95^{+0.04}_{-0.06}$ & $3.53^{+0.67}_{-0.64}$ & $1.28_{-0.74}^{+0.36}$ & $< 0.40$ & SMC & $0.49_{-0.10}^{+0.00}$ & $0.32_{-0.16}^{+0.22}$ & LMC \\
\hline
\end{tabular}
\end{minipage}
\end{table*}

\section{Conclusions}
\label{sec:cls}

In this paper we have computed rest-frame extinctions for a sample complete in redshift of GRBs \citep{Sal11}. The selection criteria of the sample rely only on observational high-energy parameters of the prompt emission and therefore our sample is not biased against dusty sight-lines. It is expected that our inferences hold for the general population of GRBs. 

Our main result is that the extinction suffered by the analyzed GRBs does not seem to follow a single distribution. 87\% of the events are absorbed by less than about 2 magnitudes, and 50\% only suffer from less than 0.3-0.4\,mag extinction. 13\% of the events are instead more absorbed, and this population of GRBs may be due to highly dusty molecular clouds/star forming regions associated with the $\gamma$-ray burst progenitor or along the afterglow line of sight, and/or to massive dusty host galaxies. The true percentage of GRB afterglows showing high absorption is probably higher, since most of the events without reliable redshift measurements are likely part of this class.

No clear evolution in the dust extinction properties is evident within the redshift range of our sample ($\sim 0.1 - 5.5$),  although the largest extinctions are at  $z\sim1.5-2$, close to the expected peak of the star formation rate \citep[$z \sim 2 - 3$, ][]{HoBe06}. Those GRBs classified as dark \citep{Mel11} show higher than average extinction, confirming past claims about the nature of under luminous optical afterglows. There is a well defined relation between optical/NIR extinction and X-ray hydrogen-equivalent column density, although with a gas-to-dust ratio well above that observed in LG environments. Dust extinction does not seem to correlate with GRB energetics or luminosity, as it could be expected if dust properties are affected by the photon flux from the high-energy event.

\section*{Acknowledgments}

This work has been supported by ASI grant I/004/11/0 and by PRIN-MIUR 2009 grants. The Dark Cosmology Centre is funded by the Danish National Research Foundation. This research has made use of the APASS database, located at the AAVSO web site. Funding for APASS has been provided by the Robert Martin Ayers Sciences Fund. S.C. thanks Cristiano Guidorzi for enlightening discussions and Paul Kuin for having provided data in a suitable format for analysis. A.G. acknowledges funding from the Slovenian Research Agency and from the Centre of Excellence for Space Science and Technologies SPACE-SI, an operation partly financed by the European Union, the European Regional Development Fund, and the Republic of Slovenia. We thank the referee, Bruce Gendre, for his useful comments.

\appendix

\section{GRBs analysed in our sample}
\label{sec:grbs}

In this section we will frequently refer to various photometric bands. With capital letters we refer to the Johnson-Morgan-Cousin bands: $U, B, V, R_{\rm c}, I_{\rm c}$. With primed small letters we refer to the SDSS photometric system $u', g', r', i', z'$\footnote{http://www.sdss.org/}. \textit{Swift}-UVOT bands are again referred with lower-case letters $uvw2,  uvm2, uvw1, u, b, v, wh$. In a few cases systematic errors to bring the fit quality to a level adequate for error modeling were introduced. Which data were selected from the literature and specific comments are reported below for each analyzed event.  

\paragraph*{GRB\,050318}

{\em Swift}-UVOT $u$, $b$ and $v$ (AB) magnitudes, reported in \citet{Sti05}, and an $R_{\rm c}$ band measurement, reported by \citet{Ber05}, are available. Calibrations of these two datasets do not appear to be in agreement, and therefore we neglected the $R_{\rm c}$ point.

\paragraph*{GRB\,050401}

We collected data from \citet{Ryk05,deP06,Wat06,Kam09} in the $V, R_{\rm c}, I_{\rm c}I_{\rm c}, J, H$ and $K$ bands. However, according to the authors, $R$ band data in \citet{deP06} are poorly calibrated and have not been used in the analyses. In addition $R$ band data in \citet{Wat06} and \citet{Kam09} are in strong disagreement. Given the more detailed calibration process reported by \citet{Kam09} we chose to use these data.

\paragraph*{GRB\,050416A}

We collected data from \citet{Hol07, Sod07} and \citet{Pri05} in the {\em Swift}-UVOT $uvm2, u, b, v$ and $R_{\rm c}, I_{\rm c}, z$ and $K$ bands. From this dataset we did not consider the GCN data due to the very preliminary calibration and one $I$ point in \citet{Hol07} at about 1150\,min that was clearly discrepant. We also removed {\em Swift}-UVOT data with error larger than 0.4\,mag.

\paragraph*{GRB\,050525A}

This event was observed intensively. We collected {\em Swift}-UVOT data in the $uvw2, w1. u, b, v$ bands from \citet{Blu06}.  We neglected all data with calibration based only on preliminary photometry.

\paragraph*{GRB\,050802}

We collected data from \citet{Oat07} in the  {\em Swift}-UVOT $uvw1, u, b, v$ bands and from Table\,\ref{tab:050802} in the $u', B, V, R_{\rm c}$ and $I_{\rm c}$ bands. In \citet{Tes05} preliminary TNG\footnote{http://www.tng.iac.es/} photometry was reported. Here we re-calibrated all the available data following the field calibration by \citet{Hen05} for $BVR_cI_c$ and SDSS\footnote{http://skyserver.sdss3.org/dr8/en/} photometry for $u'$. We neglected {\em Swift}-UVOT data obtained after about 8\,hr after the GRB and magnitudes with errors larger than 0.4\,mag. In any case {\em Swift}-UVOT still shows a considerable scatter, likely intrinsic, and to obtain a satisfactory fit we had to introduce a 15\% systematic error to be added in quadrature. 

\begin{table}
\caption{TNG data for GRB\,050802. $u'$ magnitude is in the AB system, $BVR_cI_c$ are in the Vega system.}
\label{tab:050802}
\begin{tabular}{cccccc}
\hline
$t-t_0$ & Exposure & Magnitude & Band & Telescope \\
 (min)   & (s)         & \\
\hline
 740.0 & 720 & $22.73 \pm 0.15$ & $u'$ & TNG \\
727.5 & 240 & $21.81 \pm 0.04$ & $B$ & TNG \\
722.0 & 120 & $21.37 \pm 0.05$ & $V$ & TNG \\
2135.0 & 600 & $22.85 \pm 0.07$ & $V$ & TNG \\
716.5 & 120 & $21.07 \pm 0.05$ & $R_c$ & TNG \\
709.5 & 240 & $20.44 \pm 0.06$ & $I_c$ & TNG \\
2197.0 & 360 & $21.90 \pm 0.15$ & $I_c$ & TNG \\
\hline
\end{tabular}
\end{table}

\paragraph*{GRB\,050922C}

We collected data from \citet{Oat09} in the  {\em Swift}-UVOT $u, b, v$ bands. In Table\,\ref{tab:050922C} we report VLT\footnote{http://www.eso.org/} photometry in $B, V$ and $R_{\rm c}$ bands calibrated by Landolt standard field stars and preliminarily published in \citet{Cov05}. In the fit we used only data earlier than about 20\,min. 

\begin{table}
\caption{TNG data for GRB\,050922C. Magnitudes are in the Vega system.}
\label{tab:050922C}
\begin{tabular}{cccccc}
\hline
$t-t_0$ & Exposure & Magnitude & Band & Telescope \\
 (min)   & (s)         & \\
 \hline
1700.0 & 180 & $22.41 \pm 0.03$ & $B$ & VLT \\
1686.7 & 200 & $22.00 \pm 0.02$ & $V$ & VLT \\
1692.4 & 120 & $21.52 \pm 0.04$ & $R_c$ & VLT \\
\hline
\end{tabular}
\end{table}

\paragraph*{GRB\,060206}

This event was studied by several groups \citep{Woz06,Mon06,Sta07,Cur07}. We collected optical data from \citet{Hai06b,Mon06,Cur07,Sta07}  and \citet{Oat09} in the {\em Swift}-UVOT $v$ and  in the $r', R_{\rm c}, I_{\rm c}$ bands, and NIR data from \citet{Ala06} and \citet{Ter06} in the $J, H, K$ bands.  In Table\,\ref{tab:060206} we report Asiago\footnote{http://www.pd.astro.it/asiago/} and TNG photometry in the $r'$ and $i'$ bands calibrated by SDSS\footnote{http://skyserver.sdss3.org/dr8/en/} photometry of field stars, which was preliminarily published in \citet{Mal06a}. The available data come from several different telescopes and in spite of the efforts devoted to cross-calibration by the various authors we had to introduce a 5\% systematic error to be added in quadrature. In addition, $B$ and $V$ bands are bluer than Lyman-$\alpha$ at the redshift of this event. As shown, e.g., by \citet{Aok09} this spectral range is affected by a dense Lyman-$\alpha$ forest making it difficult to model its absorption. We therefore limited our analysis to data redder than Lyman-$\alpha$.

\begin{table}
\caption{TNG data for GRB\,060206. Magnitudes are in the AB system.}
\label{tab:060206}
\begin{tabular}{cccccc}
\hline
$t-t_0$ & Exposure & Magnitude & Band & Telescope \\
 (min)   & (s)         & \\
 \hline
1158.5 & 2000 & $19.95 \pm 0.06$ & $r'$ & Asiago \\
2695.7 & 3600 & $21.02 \pm 0.06$ & $r'$ & Asiago  \\
11534.1 & 3000 & $23.52 \pm 0.35$ & $r'$ & TNG \\
1229.2 & 2700 & $19.48 \pm 0.05$ & $i'$ & Asiago \\
\hline
\end{tabular}
\end{table}

\paragraph*{GRB\,060210}

Data for this event were published by \citet{Sta07}, $R_{\rm c}$, \citet{Cur07b}, $R_{\rm c}$ and $I_{\rm c}$, and \citet{Cen09}, $R_{\rm c}, i'$ and $z'$ bands. Unfortunately, as already pointed out by \citet{Kan10}, there is a calibration offset between $I_{\rm c}$ data from \citet{Cur07b} and $i$ data from \citet{Cen09}. We had therefore to add in quadrature a 5\% systematic error.

\paragraph*{GRB\,060306}

No afterglow detection was reported for this event. Upper limits are reported by \citet{deP06b} in the \textit{Swift}-UVOT bands and by \citet{Nys06,Che06} and \citet{Lam06} in the NIR. Only a lower limit for $A_V$ could be derived. The redshift for this event was questioned by \citet{Jak12,Kru12}, and a solution at lower redshift is reported by \citet{Per13}.

\paragraph*{GRB\,060614}

This low-redshift event was widely followed. We collected \textit{Swift}-UVOT $uvw2,  uvm2, uvw1, u, b, v$ and $wh$ data from \citet{Man07} and $B, V, R_{\rm c}, I_{\rm c}, J, K$ data from \citet{Del06}. These two main datasets show small calibration inconsistencies and we had to introduce a 5\% systematic error to be added in quadrature.

\paragraph*{GRB\,060814}

The afterglow was detected in the NIR \citep{Lev06,Cen06c,Lev06b}, with its emission already contaminated by the relatively bright host galaxy. The same is true for our $r'$ and $i'$ VLT observations calibrated by SDSS photometry of field stars, which were preliminarily discussed in \citet{Mal06b} and reported in Table\,\ref{tab:060814}. We derived a lower limit on $A_V$ assuming that the afterglow cannot be brighter than these detection, likely dominated by the host galaxy contribution. With reference to Figure\,9 in \citet{Jak12} we have tried to isolate the photometry of object A only.

\begin{table}
\caption{VLT data for GRB\,060814. The host galaxy is likely to dominate these measurements. Magnitudes are in the AB system. }
\label{tab:060814}
\begin{tabular}{cccccc}
\hline
$t-t_0$ & Exposure & Magnitude & Band & Telescope \\
 (min)   & (s)         & \\
 \hline
50.1 & 900 & $24.39 \pm 0.11$ & $r'$ & VLT \\
1469.9 & 960 & $24.71 \pm 0.20$ & $r'$ & VLT \\
89.3 & 360 & $24.11 \pm 0.22$ & $i'$ & VLT \\
1464.4 & 1440 & $24.56 \pm 0.31$ & $i'$ & VLT \\
\hline
\end{tabular}
\end{table}

\paragraph*{GRB\,060904A}

No adequately calibrated observations of this afterglow have been published, and the redshift is also not known.

\paragraph*{GRB\,060908}

Data for this event are collected by \citet{Cen09} ($g', R_{\rm c}, i'$ and $ z'$ bands) and \citet{Cov10} (\textit{Swift}-UVOT $u, v$ and $wh$ and $B, V, R_{\rm c}, I_{\rm c},  J, H$ and $K$ bands). The total dataset is remarkable although due to the many different telescope/instrument combinations we had to add in quadrature an 8\% systematic error, as already discussed in \citet{Cov10}.

\paragraph*{GRB\,060912A}

Data for this event are collected from \citet{Oat09} in the \textit{Swift}-UVOT $uvw2, uvm2, uvw1, u, b, v$ and $wh$ bands and from \citet{Den09} in the $R_{\rm c}$ band. Lyman-$\alpha$ absorption affecting the bluest \textit{Swift}-UVOT filter was modeled superposing to the power-law SED a Voigt profile with parameters reported in \citet{Den09}.

\paragraph*{GRB\,060927}

Calibrated data for this event were reported by \citet{Rui07} in the $i', R_{\rm c}, I_{\rm c}$ and $K$ bands. The $R_{\rm c}$ band flux was corrected for Lyman-$\alpha$ absorption with parameters obtained by \citet{Rui07}.

\paragraph*{GRB\,061007}

Data are collected from \citet{Mun07} ($R_{\rm c}$ band) and \citet{Sch07} (\textit{Swift}-UVOT $uvw1, u, b, v$ and $wh$ bands). Cross-calibrations of the datasets required to add in quadrature a 3\% systematic error.

\paragraph*{GRB\,061021}

Data are collected from \citet{Oat09} in the \textit{Swift}-UVOT $uvw2, uvm2, uvw1, u, b, v$ and $wh$ bands.

\paragraph*{GRB\,061121}

Data are collected from \citet{Oat09} in the \textit{Swift}-UVOT $uvw1, u, b, v$ and $wh$ bands and from \citet{Kan10} in the $V, R_{\rm c}$ and $I_{\rm c}$ bands. We removed the $uvw1$ band since it is likely heavily affected by Lyman-$\alpha$ absorption.

\paragraph*{GRB\,061222A}

Only $K$ band afterglow detection was reported by \citet{Cen06a,Cen06b}. 
The available NIR data can not constrain the fit parameters and we derived the extinction by extrapolating the X-ray spectrum.

\paragraph*{GRB\,070306}

Data in the $H$ and $K$ bands are collected from \citet{Jan08}.

\paragraph*{GRB\,070328}

No afterglow detection was reported for this event and the redshift is not known. Upper limits in \textit{Swift}-UVOT bands are reported by \citet{Mar07}.

\paragraph*{GRB\,070521}

No solid afterglow detection was reported for this event. The available photometry refers to the relatively bright host galaxy. Upper limits from \citet{Min07} have been used to derive a lower limit on $A_V$.

\paragraph*{GRB\,071020}

Data are collected from \citet{Cen09} in the $R_{\rm c}, i'$ and $z'$ bands and from \citet{Blo07} in the $J$ and $H$ bands. The light curve of this event shows a considerable scatter, and we had to add in quadrature an 11\% systematic error.

\paragraph*{GRB\,071112C}

Data in the \textit{Swift}-UVOT $uvw1, u, b, v$ and $wh$ bands were collected from \citet{Oat07b}. NIR $J$ and $K$ bands data were collected from \citet{Ueh10} and $V, R_{\rm c}$ and $I_{\rm c}$ bands were collected from \citet{Hua12}. $g', r'$ and $'i'$ bands observations obtained at Asiago and TNG and calibrated following the SDSS are reported in Table\,\ref{tab:071112C}.

\begin{table}
\caption{Asiago and TNG data for GRB\,071112C. Magnitudes are in the AB system.}
\label{tab:071112C}
\begin{tabular}{cccccc}
\hline
$t-t_0$ & Exposure & Magnitude & Band & Telescope \\
 (min)   & (s)         & \\
 \hline
97.7 & 600 & $20.88 \pm 0.15$ & $g'$ & Asiago  \\
119.7 & 300 & $20.90 \pm 0.17$ & $r'$ & Asiago \\
126.7 & 300 & $21.01 \pm 0.28$ & $i'$ & Asiago \\
3274.4 & 300 & $24.04 \pm 0.09$ & $r'$ & TNG \\
32036.9 & 600 & $22.36 \pm 0.12$ & $i'$ & TNG \\
125409.9 & 600 & $23.93 \pm 0.20$ & $i'$ & TNG \\
\hline
\end{tabular}
\end{table}

\paragraph*{GRB\,071117}

The afterglow discovery was reported by \citet{Blo07b,Blo07c}. The redshift was derived by \citet{Jak07} when the host galaxy was likely already dominating the afterglow light. Data in Table\,\ref{tab:071117}, calibrated by standard stars observed with the VLT, supersede those reported in \citet{Blo07b,Blo07c} and \citet{Jak07}. These data and upper limits from \citet{BeGo07} do not allow to constrain the rest frame extinction.

\begin{table}
\caption{VLT and Gemini-S observations of GRB\,071117. Magnitudes are in the Vega system.}
\label{tab:071117}
\begin{tabular}{cccccc}
\hline
$t-t_0$ & Exposure & Magnitude & Band & Telescope \\
 (min)   & (s)         & \\
 \hline
558 & 300 & $23.04 \pm 0.10$ & $R_{\rm c}$ & VLT  \\
574 & 900 & $23.02 \pm 0.04$ &  $R_{\rm c}$ & Gemini-S  \\
694 & 540 & $23.02 \pm 0.05$ & $R_{\rm c}$ & VLT \\
2009 & 1440 & $23.53 \pm 0.04$ &  $R_{\rm c}$ & Gemini-S  \\
\hline
\end{tabular}
\end{table}

\paragraph*{GRB\,080319B}

This is the famous ``naked eye" burst. Data in the $V$ were collected from \citet{Pan09}. From \citet{Blo09} we used data in $r'$ band and in the $B, V, R_{\rm c}$ and $I_{\rm c}$ bands. Finally, data in the \textit{Swift}-UVOT $uvw2, uvm2, uvw1, u, b, v$ and $wh$ bands and in the $R_{\rm c}, I_{\rm c}$ bands were obtained from \citet{Rac08}. We however did not consider in the fit the \textit{Swift}-UVOT $uvw2, uvm2$ and $uvw1$ filters for possible Lyman-$\alpha$ contamination. In addition in many cases the reported photometric errors are likely neglecting the absolute calibration error judging from the unrealistic small quoted uncertainties. Therefore we had to add in quadrature a 3\% systematic error.

\paragraph*{GRB\,080319C}

Data are collected from \citet{Cen09} in the $R_{\rm c}, i'$ and $z'$ bands.

\paragraph*{GRB\,080413B}

Data in the $g', r', i'$ and $z'$ bands, in the $J, H$ and $K_{\rm s}$ bands and in the \textit{Swift}-UVOT $v$ band were collected from \citet{Fin11}.  We added in quadrature a 3\% systematic error to compensate for the very low photometric errors of the available data possibly neglecting absolute calibration uncertainties.

\paragraph*{GRB\,080430}

The only reliably calibrated data published so far for this event are in the \textit{Swift}-UVOT $uvw2, uvm2, uvw1, u, b, v$ and $wh$ bands from \citet{Lan08}.

\paragraph*{GRB\,080602}

No afterglow detection has been published for this event. In addition the available XRT data do not allow a meaningful comparison with available optical upper limits as in \citet{Mal08a}.

\paragraph*{GRB\,080603B}

Data in the \textit{Swift}-UVOT $uvw1, u, b$ and $v$ bands were collected from \citet{Kui08}. Data in the $g', r'$ and $i'$ bands were collected from \citet{Jel12}, and in the $J, H$ and $K$ bands from \citet{Mil08}. We did not use data bluer than the Lyman-$\alpha$ line at this GRB redshift due to strong flux suppression \citep[see, e.g.][]{Jel12}.

\paragraph*{GRB\,080605}

Data in the $g', r', i'$ and $z'$ bands and in the $J, H$ and $K_{\rm s}$ bands were collected from \citet{Zaf12}.

\paragraph*{GRB\,080607}

Data in the \textit{Swift}-UVOT $v$ band were collected from \citet{Sch08}. Data in the $V, R_{\rm c}, I_{\rm c}, J, J$ and $K_{\rm s}$ bands, and in the $i'$ and $z'$ bands were collected from \citet{Per11}. We added in quadrature a 5\% systematic error to compensate for cross-calibration problems and intrinsic short-term variability in the data.

\paragraph*{GRB\,080613B}

No optical afterglow detection has been reported.

\paragraph*{GRB\,080721}

Data in the $V$ and $R_{\rm c}$, and in the $r'$ and $i'$ bands were collected from \citet{Sta09}. We did not consider data bluer than Lyman-$\alpha$.

\paragraph*{GRB\,080804}

Data in the \textit{Swift}-UVOT $u, b, v$ and $wh$ bands were collected from \citet{Kui08b} and data from GCN circulars to define the temporal decay only.

\paragraph*{GRB\,080916A}

Data in the \textit{Swift}-UVOT $u, b, wh$ bands were collected from \citet{Oat08,Zia08}. $H$ band observations obtained with REM and calibrated following 2MASS stars in the field are reported in Table\,\ref{tab:080916A}.

\begin{table}
\caption{REM data for GRB\,080916A. Magnitudes are in the Vega system.}
\label{tab:080916A}
\begin{tabular}{cccccc}
\hline
$t-t_0$ & Exposure & Magnitude & Band & Telescope \\
 (min)   & (s)         & \\
 \hline
1.27 & 50 & $13.78 \pm 0.06$ & $H$ & REM  \\
2.59 & 50 & $14.49 \pm 0.11$ & $H$ & REM \\
3.92 & 50 & $14.12 \pm 0.08$ & $H$ & REM \\
5.26 & 50 & $14.13 \pm 0.08$ & $H$ & REM \\
6.41 & 25 & $14.37 \pm 0.14$ & $H$ & REM \\
16.25 & 100 & $15.14 \pm 0.15$ & $H$ & REM \\
\hline
\end{tabular}
\end{table}

\paragraph*{GRB\,081007}

Data in the \textit{Swift}-UVOT $u, b, v, wh$ bands, in the $B, V, R_{\rm c}, I_{\rm c}, H$ and $K_{\rm s}$ and in the $r'$ and $i'$ bands were collected from \citet{Jin12}. Due to the large number of detectors/telescopes involved we had to add in quadrature a 7\% systematic error to compensate for cross-calibration problems.

\paragraph*{GRB\,081121}

Data in the \textit{Swift}-UVOT $u, b, v$ and $wh$ bands were collected from \citet{Oat08b}.  Data in the $g', r', i', z', J, H$ and $K_{\rm s}$ were collected from \citet{Loe08}. 

\paragraph*{GRB\,081203A}

Data in the \textit{Swift}-UVOT $uvw2, uvw1, u, b, v$ and $wh$ bands were collected from \citet{deP08} and \citet{Kui09}. $R_{\rm c}$ data were collected from \citet{Fat08}. We did not consider data bluer than Lyman-$\alpha$.

\paragraph*{GRB\,081221}

No afterglow detection has been reported for this event. The upper limit reported by \citet{Mal08} was used to derive a lower limit for $A_V$.

\paragraph*{GRB\,081222}

Data in the \textit{Swift}-UVOT $u, b, v$ and $wh$ bands were collected from \citet{Bre08}. Data in the $J, H$ and $K_{\rm s}$ bands from \citet{Upd08}. REM photometry is reported in Table\,\ref{tab:081222}. These data update and substitute those reported in \citet{Cov08b}. NIR data were calibrated with 2MASS stars in the field and optical data with stars reported in the APASS catalogue\footnote{http://www.aavso.org/apass}. We did not consider data bluer than Lyman-$\alpha$. We had to add in quadrature a 7.5\% systematic error to compensate the unrealistic low errors reported in some of the available photometry possibly neglecting absolute calibration uncertainties.

\begin{table}
\caption{REM data for GRB\,081222. $J,H$ and $K_{\rm s}$ magnitudes are in the Vega system, $r'$ magnitudes in the AB system.}
\label{tab:081222}
\begin{tabular}{cccccc}
\hline
$t-t_0$ & Exposure & Magnitude & Band & Telescope \\
 (min)   & (s)         & \\
 \hline
1.37 & 50 & $11.24 \pm 0.09$ & $H$ & REM  \\
2.69 & 50 & $12.24 \pm 0.10$ & $H$ & REM  \\
4.02 & 50 & $12.95 \pm 0.07$ & $H$ & REM  \\
10.01 & 30 & $15.94 \pm 0.11$ & $r'$ & REM  \\
10.67 & 30 & $16.00 \pm 0.11$ & $r'$ & REM  \\
11.32 & 30 & $16.10 \pm 0.12$ & $r'$ & REM  \\
11.98 & 30 & $16.28 \pm 0.13$ & $r'$ & REM  \\
12.61 & 150 & $14.57 \pm 0.07$ & $H$ & REM  \\
12.63 & 30 & $16.28 \pm 0.14$ & $r'$ & REM  \\
12.29 & 30 & $16.30 \pm 0.14$ & $r'$ & REM  \\
13.94 & 30 & $16.36 \pm 0.15$ & $r'$ & REM  \\
14.59 & 30 & $16.20 \pm 0.16$ & $r'$ & REM  \\
17.27 & 30 & $16.51 \pm 0.17$ & $r'$ & REM  \\
17.92 & 30 & $16.68 \pm 0.19$ & $r'$ & REM  \\
18.58 & 30 & $16.62 \pm 0.18$ & $r'$ & REM  \\
19.23 & 30 & $16.73 \pm 0.20$ & $r'$ & REM  \\
19.89 & 30 & $16.63 \pm 0.18$ & $r'$ & REM  \\
28.08 & 150 & $15.93 \pm 0.15$ & $J$ & REM  \\
\hline
\end{tabular}
\end{table}

\paragraph*{GRB\,090102}

Data in the \textit{Swift}-UVOT $u, b, v$ and $wh$ bands were collected from \citet{Cur09}. Data in the $g', R_{\rm c}, i', z', J, H$ and $K_{\rm s}$ were collected from \citet{Gen10}.

\paragraph*{GRB\,090201}

The afterglow detection, likely already substantially contaminated by the host galaxy, was reported by \citet{Dav09}. Data reported in \citet{Dav09} are superseded by those reported in Table\,\ref{tab:090201}. Optical data were calibrated by observation of standard star frames and NIR data following the 2MASS catalogue. The redshift for this event is reported in \citet{Kru13b}.

\begin{table}
\caption{VLT data for GRB\,090201. Magnitudes are in the Vega system.}
\label{tab:090201}
\begin{tabular}{cccccc}
\hline
$t-t_0$ & Exposure & Magnitude & Band & Telescope \\
 (min)   & (s)         & \\
 \hline
434.0 & 720 & $24.12 \pm 0.08$ & $R_{\rm c}$ & VLT  \\
916574 & 5000 & $24.40 \pm 0.09$ & $R_{\rm c}$ & NTT  \\
451.0 & 480 & $23.75 \pm 0.11$ & $I_{\rm c}$ & VLT \\
554.0 & 1200 & $22.05 \pm 0.13$ & $J$ & VLT \\
580.4 & 1200 & $21.31 \pm 0.15$ & $H$ & VLT \\
608.0 & 1200 & $20.19 \pm 0.14$ & $K_{\rm s}$ & VLT \\
\hline
\end{tabular}
\end{table}

\paragraph*{GRB\,090424}

Data in the \textit{Swift}-UVOT $u, b, v, wh$ bands and in the $R_{\rm c}$ and $I_{\rm c}$ were collected from \citet{Jin12}. Data in the $g', r', i', z', J, H$ and $K_{\rm s}$ were collected from \citet{Oli09}. Data in the $V, I_{\rm c}, J$ and $K_{\rm s}$ bands from \citet{Cob09}. Due to the large number of detectors/telescopes involved we had to add in quadrature a 7.5\% systematic error to compensate for cross-calibration problems.

\paragraph*{GRB\,090709A}

Data in the $z', J, H$ and $K_{\rm s}$ bands were collected from \citet{Cen10}. The redshift for this event is reported by \citet{Per13}. Data show a considerable scatter, and we had to add in quadrature a 9\% systematic error.

\paragraph*{GRB\,090715B}

Data in the $R_{\rm c}$ and $I_{\rm c}$ bands were collected from \citet{Hai09}, data in the $r', i'$ and $z'$ from Virgili et al. (2013, in preparation).

\paragraph*{GRB\,090812}

Data in the \textit{Swift}-UVOT $u$ and $b$ bands were collected from \citet{Sch09}. Data in the $g', r', i', z'$ and $j$ bands were collected from \citet{Upd09a} and Virgili et al. (2013, in preparation). We did not use data bluer than Lyman-$\alpha$.

\paragraph*{GRB\,090926B}

No reliably calibrated data have been published for this event although in \citet{Gre11} a SED analysis was reported. Upper limits from \citet{Xu09} did not allow to constrain rest frame extinction.

\paragraph*{GRB\,091018}

Data in the \textit{Swift}-UVOT $uvw1, u, b, v, wh$ bands, in the $g', r', i', z'$ bands, and in the $J, H$ and $K$ bands were collected from \citet{Wie12}.  Due to the large number of detectors/telescopes involved we had to add in quadrature a 2.5\% systematic error to compensate for cross-calibration problems.

\paragraph*{GRB\,091020}

Data in the \textit{Swift}-UVOT $uvw1, u, b, v$ and $wh$ bands were collected from \citet{Oat09b} and \citet{Rac09} and in the $B, V, R_{\rm c}$ and $I_{\rm c}$ from \citet{Kan09,Kan09b}. We did not use data bluer than Lyman-$\alpha$.

\paragraph*{GRB\,091127}

Data in the $g', r', i', z', J$ and $H$ bands were collected from \citet{Fil11}, data in the $I_{\rm c}$ band from \citet{Verg11}, data in the \textit{Swift}-UVOT $uvw1$ and $u$ bands, and in the $B, V, R_{\rm c}$ and $I_{\rm c}$ bands from \citet{Cob10}. Due to the large number of detectors/telescopes involved we had to add in quadrature a 2\% systematic error to compensate for cross-calibration problems.

\paragraph*{GRB\,091208B}

Data in the \textit{Swift}-UVOT $u$ and $b$ bands were collected from \citet{deP09}, data in the $g', r', i', z', J$ and $H$ bands were collected from \citet{Upd09}. We however did not use data bluer than Lyman-$\alpha$ and collected not-properly calibrated data from GCNs to constrain the temporal decay only.

\paragraph*{GRB\,100615A}

No detection is reported for this afterglow. Upper limits are discussed in \citet{DeSt11}. Redshift was recently measured by \citet{Kru13}.

\paragraph*{GRB\,100621A}

Data in the $g', r', i', z', J, H$ and $K_{\rm s}$ bands were collected from \citet{Kru11} and data in the $J, H$ and $K_{\rm s}$ were collected from \citet{Nai10}. Data from \citet{Kru11} are late enough to be considerably affected by the host galaxy brightness, in particular in the bluest available bands. We deal with this uncertainty by adding in quadrature a 7.5\% systematic error.

\paragraph*{GRB\,100728B}

Data in the \textit{Swift}-UVOT $u, b$ and $wh$ bands were collected from \citet{Oat10}, data in the $R_{\rm c}$ band from \citet{Vol10} and data in the $J, H$ and $K_{\rm s}$ bands from \citet{Oli10}.

\paragraph*{GRB\,110205A}

Data in the \textit{Swift}-UVOT $u, b, v$ and $wh$ bands, in the $g', r', i'$ and $z'$ bands and in the $J, H$ and $K_{\rm s}$ bands were collected from \citet{Cuc11}. Data in the $U, B, V, R_{\rm c}, I_{\rm c}, g', r', i'$ and $z'$ bands from \citet{Zhe12}. Data in the $B, V, R_{\rm c}$ and $I_{\rm c}$ bands from \citet{Gen12}. Data in the $g', R_{\rm c}$ and $I_{\rm c}$ bands were collected from \citet{Kur11,Kur11b}. Finally, data in the $r'$ band were collected from \citet{Ura11}. To compensate for cross-calibration problems in different datasets we added a 2\% systematic error. We did not use data bluer than Lyman-$\alpha$.

\paragraph*{GRB\,110503A}

Data in the \textit{Swift}-UVOT $uvw1, uvm2, uvw2, u, b, v$ and $wh$ bands were collected from \citet{Oat11} and data in the $J$ and $H$ bands were collected from \citet{MoBl11}. To compensate for cross-calibration problems in preliminary datasets we added a 7.5\% systematic error. We did not use data bluer than Lyman-$\alpha$.

\begin{table*}
\caption{Results of the analysis performed in this work for the GRBs in the sample. We report the spectral index and the amount of rest-frame reddening for each applied extinction recipe: MW, LMC and SMC. Spectral slopes and extinctions are computed leaving the spectral slope free or imposing a prior from X-ray data as described in Sect.\,\ref{sec:dtmth}.}
\label{tab:allres}
\begin{small}
\begin{tabular}{lclccclcccccc}
\hline
Event                & $t_{\rm min}, t_{\rm max}$  & Ext. curve & \multicolumn{3}{c}{No X-ray prior} &  \multicolumn{3}{c}{With X-ray prior} \\
                          &   (min)                                     &                    & $\beta_{\rm o}$                         & $E_{B-V}$                 & $\chi^2/$dof  & $\beta_{\rm o}$ & $E_{B-V}$ & $\chi^2/$dof  \\
\hline
GRB\,050318 & 43, 860                                   & MW            & $4.78_{-1.00}^{+1.11}$ & $0.27_{-0.19}^{+0.20}$ & 1.00 (4.00/4) & 1.02 & 0.00 & 16.25 (65.01/4) \\
                          &                                                 & LMC           & $2.23_{-1.18}^{+1.14}$ & $0.34_{-0.18}^{+0.29}$ & 1.00 (4.00/4) & $1.02_{-0.13}^{+0.00}$ & $0.57^{+0.17}_{-0.15}$ & 1.47 (5.86/4) \\
                          &                                                 & SMC           & $-2.00_{-4.90}^{+3.89}$ & $0.50_{-0.35}^{+0.40}$ & 1.00 (4.00/4)  & $0.89^{+0.13}_{-0.00}$ & $0.25^{+0.06}_{-0.06}$ & 1.25 (5.02/4) \\
\hline
GRB\,050401  & 29, 1400                               & MW             &  0.70                                  &  0.00                    & 2.16 (12.96/6) & +0.48                              & 0.03                                   & 2.41 (14.43/6) \\
		        &                                                & LMC            & -0.41                                    & 0.20                    & 1.84 (11.04/6) & +0.19                               & 0.11                                   & 2.12 (12.71/6) \\
                           &                                                & SMC            & $-0.42_{-0.73}^{+0.65}$ & $0.16_{-0.09}^{+0.10}$ & 1.43 (8.59/6)   & $+0.19^{+0.29}_{-0.00}$& $0.07^{+0.02}_{-0.04}$ & 1.64 (9.85/6) \\
\hline
GRB\,050416A & 43, 10000 			   & MW               & $0.37^{+0.78}_{-1.03}$ &  $0.34_{-0.27}^{+0.35}$ & 0.84 (20.17/24) & $0.48^{+0.24}_{-0.01}$& $0.30^{+0.04}_{-0.12}$ & 0.84 (20.20/24) \\
			& 				            & LMC              & $0.87^{+0.39}_{-0.47}$ & $0.14_{-0.10}^{+0.13}$  & 0.78 (18.73/24) & $0.72^{+0.00}_{-0.25}$& $0.18^{+0.09}_{-0.03}$ & 0.79 (18.98/24) \\
			& 					   & SMC             & $1.04^{+0.25}_{-0.33}$ & $0.08_{-0.04}^{+0.08}$  & 0.78 (18.60/24) & $0.72_{-0.17}^{+0.00}$ & $0.16^{+0.05}_{-0.03}$ & 0.86 (20.73/24) \\         
\hline
GRB\,050525A & 2.9, 7.8                                & MW                & 1.48                                   & 0.00                                                      & 2.16 (38.92/18) & 0.73 & 0.17 & 3.02 (54.36/18) \\                        
			&					   & LMC 		 & 0.73                                   & 0.10                                                     & 1.81 (32.63/18) & 0.73 & 0.10 & 1.82 (32.84/18) \\
			&                                               & SMC              & $0.27^{+0.48}_{-0.52}$ & $0.12^{+0.05}_{-0.04}$                   & 1.42 (25.63/18) & $0.45_{-0.00}^{+0.28}$ & $0.10^{+0.00}_{-0.03}$ & 1.45 (26.06/18) \\
\hline
GRB\,050802 & 86, 2900                                & MW                & $1.78_{-0.42}^{+0.31}$ & $< 0.16 (0.01)$                   & 1.71 (10.47/6) & $0.43$ & $0.22$ & 4.70 (28.23/6) \\
		       &					   & LMC              & $0.79_{-1.24}^{+1.08}$ & $< 0.55 (0.13)$                   & 1.51 (9.05/6) & $0.43_{-0.11}^{+0.00}$ & $0.17_{-0.03}^{+0.05}$   & 1.53 (9.21/6)\\
		       &					   & SMC              & $0.34_{-1.42}^{+1.10}$ & $< 0.43 (0.13)$                   & 1.31 (7.85/6) & $0.34_{-0.02}^{+0.09}$ & $0.13_{-0.03}^{+0.03}$   & 1.31 (7.85/6)\\
\hline
GRB\,050922C &  3.9, 16                               & MW                 & $1.21^{+0.25}_{-0.65}$ & $< 0.09 (0.01)$ & 1.20 (19.22/16) & $0.81^{+0.00}_{-0.13}$ & $0.05_{-0.01}^{+0.02}$ & 1.25 (19.92/16)\\
			&					  & LMC               & $0.79^{+0.68}_{-2.02}$ & $< 0.22 (0.04)$ & 1.20 (19.21/16) & $0.78^{+0.03}_{-0.10}$ & $0.04_{-0.01}^{+0.01}$ & 1.20 (19.21/16) \\
			&					  & SMC               & $-0.30^{+1.47}_{-4.74}$ & $< 0.56 (0.10)$ & 1.20 (19.21/16)  & $0.68^{+0.13}_{-0.00}$ & $0.03_{-0.01}^{+0.01}$ & 1.20 (19.21/16)\\ 
\hline
GRB\,060206 &  72, 330                              & MW 		  & $0.85^{+0.07}_{-0.33}$ & $< 0.13 (0.00)$ & 1.76 (8.79/5) & $0.85^{+0.07}_{-0.33}$ & $< 0.13 (0.00)$ & 1.76 (8.79/5)  \\
		       &					  & LMC 		  & $0.02^{+0.64}_{-0.60}$ & $0.11_{-0.08}^{+0.08}$ & 1.14 (5.72/5) & $0.27^{+0.50}_{-0.00}$ & $0.08_{-0.07}^{+0.01}$ & 1.20 (6.01/5)\\
		       &					  & SMC 		  & $0.46^{+0.41}_{-0.39}$ & $< 0.09 (0.04)$ & 1.39 (6.93/5) & $0.46^{+0.41}_{-0.39}$ & $< 0.06 (0.04)$ & 1.39 (6.93/5)\\		
\hline
GRB\,060210 & 8.6, 43				 & MW 		& $4.47^{+0.28}_{-1.68}$ & $< 1.00 (0.00)$ & 1.06 (22.29./21) & $1.13^{+0.00}_{-0.10}$ & $0.60_{-0.07}^{+0.10}$ & 1.24 (26.14/21)\\
			&					 & LMC 		& $4.47^{+0.28}_{-4.86}$ & $< 0.96 (0.00)$ & 1.06 (22.29/21) & $1.13^{+0.00}_{-0.10}$ & $0.28_{-0.04}^{+0.05}$ & 1.13 (23.77/21)\\
			&					 & SMC 		& $4.47^{+0.28}_{-9.51}$ & $< 0.71 (0.00)$ & 1.06 (22.29/21)	& $1.13^{+0.00}_{-0.10}$ & $0.20_{-0.02}^{+0.03}$ & 1.10 (23.01/21) \\	
\hline 
GRB\,060306 &  	96			 & MW & & & & 0.86 & $> 2.8$ & \\
			&				 & LMC & & & & 0.86 & $> 2.7$ & \\
			&				 & SMC & & & & 0.86 & $> 3.0$ & \\	
\hline 
GRB\,060614 &  2300,5300		 & MW 		 & $0.07$ & $0.33$ & 2.04 (18.40/9) & $0.35$ & $0.24$ & 2.38 (21.43/9)\\
			&					 & LMC 	 & $0.03^{+0.23}_{-0.26}$ & $0.31_{-0.07}^{+0.07}$ & 1.39 (12.48/9) & $0.35$ & $0.22$ & 1.84 (16.90/9) \\
			&					 & SMC 	 & $0.18^{+0.20}_{-0.23}$ & $0.26_{-0.06}^{+0.07}$ & 1.36 (12.28/9) & $0.35^{+0.09}_{-0.00}$ & $0.22_{-0.02}^{+0.02}$ & 1.52 (13.66/9)  \\	
\hline 
GRB\,060814 &  	50			 & MW & & & & 0.56 & $> 0.4$ &		\\
			&					 & LMC & & & & 0.56 & $> 0.4$ &	 	\\
			&					 & SMC & & & & 0.56 & $> 0.4$ &	 	 \\	
\hline 
GRB\,060904A &  				 & MW 		\\
			&					 & LMC 	 	\\
			&					 & SMC 	 	 \\	
\hline 
GRB\,060908 &  	8.6, 290			 & MW 	& $0.38^{+0.06}_{-0.07}$ & $< 0.01 (0.00)$ & 0.99 (87.80/89) & $0.56^{+0.02}_{-0.00}$ & $< 0.01 (0.00)$ & 1.14 (101.60/89)	\\
			&					 & LMC 	  & $0.38^{+0.06}_{-0.17}$ & $< 0.04 (0.00)$ & 0.99 (87.80/89) & $0.56^{+0.02}_{-0.00}$ & $< 0.01 (0.00)$ & 1.14 (101.60/89)	\\
			&					 & SMC 	 & $0.22^{+0.19}_{-0.21}$ & $< 0.07 (0.02)$ & 0.97 (86.50/89)	& $0.56^{+0.02}_{-0.00}$ & $< 0.01 (0.00)$ & 1.14 (101.60/89)	 \\	
\hline 
GRB\,060912A &  	8.6, 290			 & MW 		& $1.07^{+0.15}_{-0.64}$ & $< 0.41 (0.00)$ & 0.87 (16.44/19)  & $0.90^{+0.00}_{-0.44}$ & $< 0.22 (0.03)$ & 0.89 (16.90/19)\\
			&					 & LMC 	 & $0.25^{+0.97}_{-1.60}$ & $< 0.53 (0.13)$ & 0.83 (15.77/19) & $0.46^{+0.44}_{-0.00}$ & $< 0.16 (0.10)$ & 0.83 (15.82/19)	\\
			&					 & SMC 	 & $-0.08^{+1.23}_{-1.42}$ & $< 0.41 (0.14)$ & 0.77 (14.71/19)	 & $0.46^{+0.44}_{-0.00}$ & $0.07_{-0.06}^{+0.02}$ & 0.79 (15.07/19) \\	
\hline 
GRB\,060927 &  14, 4500			 & MW & $0.75^{+0.41}_{-0.54}$ & $< 0.17 (0.00)$ & 1.75 (6.98/4) & $0.75^{+0.41}_{-0.04}$ & $< 0.10 (0.00)$ & 1.75 (6.98/4) 		\\
			&					 & LMC & $0.75^{+0.41}_{-1.45}$ & $< 0.41 (0.00)$ & 1.75 (6.98/4) & $0.75^{+0.41}_{-0.04}$ & $< 0.08 (0.00)$ & 1.75 (6.98/4) 	 	\\
			&					 & SMC & $0.75^{+0.41}_{-2.35}$ & $< 0.47 (0.00)$ & 1.75 (6.98/4) & $0.75^{+0.41}_{-0.04}$ & $< 0.06 (0.00)$ & 1.75 (6.98/4)  \\			
\hline 
\end{tabular}
\end{small}
\end{table*}

\begin{table*}
\contcaption{}
\begin{small}
\begin{tabular}{lclccclcccccc}
\hline

GRB\,061007 &  	7.2, 22			 & MW & $2.36$ & $0.13$ & 1.80 (117.23/65)	& $1.10$ & $0.37$ & 11.05 (718.29/65)	\\
			&					 & LMC & $0.87^{+0.27}_{-0.24}$ & $0.24_{-0.04}^{+0.03}$ & 1.19 (77.40/65)	 & $0.97^{+0.13}_{-0.02}$ & $0.23_{-0.01}^{+0.01}$ & 1.20 (77.76/65)   \\
			&					 & SMC & $-0.30^{+0.40}_{-0.42}$ & $0.24_{-0.03}^{+0.03}$ & 1.18 (76.94/65)	 & $0.95$ & $0.14$ & 1.44 (93.84/65)  \\	
\hline
GRB\,061021 &  	58, 360			 & MW 	& $0.40^{+0.09}_{-0.28}$ & $< 0.14 (0.00)$ & 0.45 (7.65/17) & $0.45^{+0.09}_{-0.00}$ & $< 0.04 (0.00)$ & 0.47 (7.93/17)	\\
			&					 & LMC 	& $-0.31^{+0.80}_{-1.03}$ & $< 0.35 (0.11)$ & 0.38 (6.52/17) & $0.45^{+0.09}_{-0.00}$ & $< 0.02 (0.00)$ & 0.47 (7.93/17)	  	\\
			&					 & SMC 	 & $0.04^{+0.45}_{-0.60}$ & $< 0.15 (0.04)$ & 0.40 (6.88/17) & $0.45^{+0.09}_{-0.00}$ & $< 0.02 (0.00)$ & 0.47 (7.93/17)	 \\	
\hline 
GRB\,061121 &  	50, 1300			 & MW & $0.72_{-0.20}^{+0.21}$ & $< 0.10 (0.00)$ & 0.96 (15.34/16)	& $0.47_{-0.12}^{+0.00}$ & $< 0.13 (0.01)$ & 1.11 (17.75/16)	\\
			&					 & LMC & $-0.38_{-2.52}^{+1.24}$ & $< 0.69 (0.18)$ & 0.93 (14.82/16) & $0.35_{-0.00}^{+0.12}$ & $0.06_{-0.05}^{+0.03}$ & 0.94 (15.06/16)	 	\\
			&					 & SMC 	& $-1.15^{+1.21}_{-1.41}$ & $0.19_{-0.13}^{+0.14}$ & 0.72 (11.55/16)  & $0.35_{-0.00}^{+0.12}$ & $0.04_{-0.03}^{+0.02}$ & 0.87 (13.96/16) \\	
\hline 
GRB\,061222A &  	126			 & MW & & & & $0.95^{+0.07}_{-0.06}$ & $0.90^{+0.30}_{-0.30}$ &		\\
			&					 & LMC & & & & $0.95^{+0.07}_{-0.06}$ & $0.90^{+0.40}_{-0.30}$ & \\
			&					 & SMC & & & & $0.95^{+0.07}_{-0.06}$ & $1.00^{+0.40}_{-0.30}$ & \\	
\hline 
GRB\,070306 &  	1400, 5800			 & MW 		& & & & $0.39_{-0.00}^{+0.13}$ & $1.95_{-0.44}^{+0.52}$ & $\infty$ (1.40/0)  \\
			&					 & LMC 	& & & & $0.39_{-0.00}^{+0.13}$ & $1.85_{-0.46}^{+0.50}$ & $\infty$ (1.40/0) 	\\
			&					 & SMC & & & & $0.39_{-0.00}^{+0.13}$ & $2.02_{-0.51}^{+0.52}$ & $\infty$ (1.40/0)	 	 \\	
\hline 
GRB\,070328 &  				 & MW 		\\
			&					 & LMC 	 	\\
			&					 & SMC 	 	 \\	
\hline 
GRB\,070521 &  	90			 & MW & & & & 0.40 & $> 0.8$ & \\
			&					 & LMC & & & & 0.40 & $> 0.7$ & \\
			&					 & SMC & & & & 0.40 & $> 0.8$ &  \\	
\hline 
GRB\,071020 &  	72, 220			 & MW 	$0.58^{+0.28}_{-1.05}$ & $< 0.26 (0.00)$ & 1.40 (20.97/15)	& $0.75^{+0.20}_{-0.00}$ & $< 0.04 (0.00)$ & 1.44 (21.54/15)  \\
			&					 & LMC & $0.58^{+0.28}_{-1.38}$ & $< 0.36 (0.00)$ & 1.40 (20.97/15) & $0.75^{+0.20}_{-0.00}$ & $< 0.05 (0.00)$ & 1.44 (21.54/15) 	 	\\
			&					 & SMC & $0.58^{+0.28}_{-1.71}$ & $< 0.40 (0.00)$ & 1.40 (20.97/15) & $0.75^{+0.20}_{-0.00}$ & $< 0.05 (0.00)$ & 1.44 (21.54/15) 	 	 \\	
\hline 
GRB\,071112C &  	5.8,22			 & MW 	& $0.48^{+0.19}_{-0.40}$ & $< 0.17 (0.00)$ & 1.03 (33.86/33) & $0.49^{+0.01}_{-0.40}$ & $< 0.17 (0.00)$ & 1.03 (33.86/33)	\\
			&					 & LMC 	& $0.48^{+0.19}_{-0.44}$ & $< 0.13 (0.00)$ & 1.03 (33.86/33) &  $0.49^{+0.01}_{-0.44}$ & $< 0.13 (0.00)$ & 1.03 (33.86/33)	\\
			&					 & SMC & $0.48^{+0.19}_{-0.37}$ & $< 0.10 (0.00)$ & 1.03 (33.86/33) &	 $0.49^{+0.01}_{-0.37}$ & $< 0.10 (0.00)$ & 1.03 (33.86/33)  \\	
\hline 
GRB\,071117 &  				 & MW 		\\
			&					 & LMC 	 	\\
			&					 & SMC 	 	 \\	
\hline 
GRB\,080319B &  	72, 860			 & MW 	& $0.39^{+0.02}_{-0.02}$ & $< 0.01 (0.00)$ & 1.12 (110.87/99) & $0.38^{+0.00}_{-0.02}$ & $< 0.01 (0.00)$ & 1.12 (110.98/99)	\\
			&					 & LMC 	 & $0.38^{+0.02}_{-0.04}$ & $< 0.02 (0.00)$ & 1.12 (110.87/99) & $0.38^{+0.00}_{-0.04}$ & $< 0.02 (0.00)$ & 1.12 (110.98/99)	\\
			&					 & SMC 	 & $0.22^{+0.15}_{-0.15}$ & $0.03_{-0.02}^{+0.02}$ & 1.09 (108.32/99) & $0.26^{+0.11}_{-0.00}$ & $< 0.02 (0.02)$ & 1.10 (108.46/99) \\	
\hline 
GRB\,080319C &  		18, 36		 & MW 	& $1.61^{+0.62}_{-0.68}$ & $0.06_{-0.05}^{+0.05}$ & 0.89 (1.79/2) & $0.75$ & $0.13$ & 2.52 (5.04/2)	\\
			&					 & LMC 	 & $1.37^{+0.86}_{-0.88}$ & $0.11_{-0.09}^{+0.09}$ & 0.89 (1.79/2)  & $0.75^{+0.00}_{-0.49}$ & $0.17_{-0.02}^{+0.05}$ & 1.41 (2.82/2) 	\\
			&					 & SMC 	 & $0.48^{+1.66}_{-1.64}$ & $0.26_{-0.23}^{+0.23}$ & 0.89 (1.79/2)   & $0.48^{+0.27}_{-0.24}$ & $0.26_{-0.07}^{+0.06}$ & 0.89 (1.79/2) 	 \\	
\hline 
GRB\,080413B &  	19,100			 & MW   & $0.23^{+0.04}_{-0.05}$ & $0.02_{-0.01}^{+0.01}$ & 0.92 (140.08/153) & $0.40^{+0.01}_{-0.00}$ & $< 0.01 (0.00)$ & 1.09 (166.49/153)	\\
			&					 & LMC & $0.19^{+0.07}_{-0.07}$ & $0.03_{-0.01}^{+0.01}$ & 0.91 (139.90/153) & $0.40^{+0.01}_{-0.00}$ & $< 0.01 (0.00)$ & 1.09 (166.49/153)	 \\
			&					 & SMC & $0.13^{+0.08}_{-0.08}$ & $0.04_{-0.01}^{+0.01}$ & 0.92 (140.28/153)  & $0.40^{+0.01}_{-0.00}$ & $< 0.01 (0.00)$ & 1.09 (166.49/153) \\	
\hline 
GRB\,080430 &  	1.4, 180			 & MW & $0.69^{+0.60}_{-0.55}$ & $0.12_{-0.11}^{+0.12}$ & 1.30 (5.19/4) & $0.50^{+0.12}_{-0.06}$ & $0.14_{-0.05}^{+0.04}$ & 1.68 (6.72/4) \\
			&					 & LMC & $0.50^{+0.95}_{-1.27}$ & $< 0.28 (0.10)$ & 1.65 (6.62/4) & $0.49^{+0.13}_{-0.00}$ & $0.10^{+0.02}_{-0.04}$ & 1.65 (6.62/4) \\
			&					 & SMC & $1.12^{+0.37}_{-1.00}$ & $< 0.15 (0.02)$ & 1.86 (7.45/4)  & $0.62$ & $0.06$ & 1.98 (7.94/4)\\	
\hline 
GRB\,080602  &  				 & MW 		\\
			&					 & LMC 	 	\\
			&					 & SMC 	 	 \\	
\hline 
GRB\,080603B  &  	160, 720			 & MW  & $0.52^{+0.25}_{-0.30}$ & $< 0.06 (0.00)$ & 0.41 (1.22/3)	& $0.54^{+0.09}_{-0.33}$ & $< 0.06 (0.00)$ & 0.41 (1.23/3)	\\
			&					 & LMC & $0.52^{+0.25}_{-0.67}$ & $< 0.35 (0.00)$ & 0.41 (1.22/3)	 & $0.54^{+0.09}_{-0.38}$ & $< 0.20 (0.00)$ & 0.41 (1.23/3)	\\
			&					 & SMC & $0.22^{+0.55}_{-0.74}$ & $< 0.16 (0.05)$ & 0.26 (0.77/3)	 & $0.19^{+0.44}_{-0.03}$ & $< 0.06 (0.05)$ & 0.26 (0.77/3)	 \\	
\hline 
GRB\,080605  &  	72,220			 & MW  & $1.57$ & $0.00$ & 3.42 (30.81/9) & $0.97$ & $0.12$ & 2.86 (88.34/11) \\
			&					 & LMC & $0.98$ & $0.12$ & 2.52 (22.69/9) & $0.97$ & $0.12$ & 2.52 (22.69/9) \\
			&					 & SMC & $0.65^{+0.24}_{-0.22}$ & $0.16_{-0.04}^{+0.04}$ & 0.28 (2.49/9) & $0.74^{+0.20}_{-0.04}$ & $0.14_{-0.03}^{+0.01}$ & 0.30 (2.73/9)  \\	
\hline 
\end{tabular}
\end{small}
\end{table*}

\begin{table*}
\contcaption{}
\begin{small}
\begin{tabular}{lclccclcccccc}
\hline 
GRB\,080607  &  	13, 22			 & MW & $1.73^{+0.27}_{-0.29}$ & $0.30^{+0.09}_{-0.07}$ & 0.87 (7.81/9) & $1.19^{+0.00}_{-0.11}$ & $0.43^{+0.03}_{-0.02}$ & 1.46 (13.14/9) \\
			&					 & LMC & $1.28$ & $0.31$ & 4.50 (40.49/9) & $1.19$ & $0.33$ & 4.50 (40.53/9)	 \\
			&					 & SMC & $2.94$ & $0.00$ & 5.82 (52.37/9) & $1.19$ & $0.25$ & 11.96 (107.67/9)  \\	
\hline 
GRB\,080613B  &  				          & MW 		\\
			&					 & LMC 	 	\\
			&					 & SMC 	 	 \\	
\hline 
GRB\,080721  &  	580, 58000		 & MW & $1.34^{+1.09}_{-0.99}$ & $< 0.43 (0.14)$ & 0.43 (3.90/9) & $0.96^{+0.00}_{-0.10}$ & $< 0.40 (0.11)$ & 0.46 (4.16/9) \\
			&					 & LMC & $-0.23^{+1.57}_{-2.10}$ & $< 0.60 (0.21)$ & 0.42 (3.74/9) &  $0.86^{+0.10}_{-0.00}$ & $< 0.26 (0.07)$ & 0.52 (4.65/9) \\
			&					 & SMC & $-4.13^{+5.13}_{-7.30}$ & $< 1.05 (0.42)$ & 0.39 (3.48/9) & $0.86^{+0.10}_{-0.00}$ & $< 0.12 (0.01)$ & 0.57 (5.10/9) \\	
\hline 
GRB\,080804  &  	5.8, 72			 & MW  & $0.59^{+2.30}_{-1.06}$ & $< 0.53 (0.09)$ & 0.49 (1.97/4)	&  $0.55^{+0.04}_{-0.19}$ & $0.09^{+0.08}_{-0.05}$ & 0.49 (1.97/4)	\\
			&					 & LMC & $-1.74^{+3.39}_{-10.08}$ & $< 1.51 (0.25)$ & 0.49 (1.97/4) &  $0.36^{+0.23}_{-0.01}$ & $0.08^{+0.04}_{-0.05}$ & 0.51 (2.03/4) \\
			&					 & SMC & $-8.45^{+10.10}_{-18.18}$ & $< 1.76 (0.59)$ & 0.49 (1.97/4) &	 $0.37^{+0.22}_{-0.02}$ & $0.06^{+0.03}_{-0.04}$ & 0.52 (2.07/4) \\	
\hline 
GRB\,080916A &  	5, 290			 & MW & $1.47^{+0.09}_{-0.34}$ & $< 0.22 (0.00)$ & 1.06 (7.44/7)	 &  $1.20^{+0.00}_{-0.26}$ & $0.07^{+0.08}_{-0.02}$ & 1.30 (9.07/7)	\\
			&					 & LMC & $1.47^{+0.09}_{-0.43}$ & $< 0.26 (0.00)$ & 1.06 (7.44/7)	 &  $1.20^{+0.00}_{-0.31}$ & $0.07^{+0.09}_{-0.02}$ & 1.24 (8.65/7) \\
			&					 & SMC & $1.47^{+0.09}_{-0.43}$ & $< 0.26 (0.00)$ & 1.06 (7.44/7)	 & $1.20^{+0.00}_{-0.31}$ & $0.07^{+0.09}_{-0.02}$ & 1.24 (8.67/7)	 \\	
\hline 
GRB\,081007 &  	10, 120			 & MW 	& $0.36^{+0.28}_{-0.33}$ & $0.13^{+0.09}_{-0.07}$ & 0.96 (33.38/37) & $0.86^{+0.06}_{-0.00}$ & $< 0.04 (0.01)$ & 1.13 (41.97/37	\\
			&					 & LMC 	& $0.31^{+0.37}_{-0.38}$ & $0.14^{+0.10}_{-0.09}$ & 0.99 (36.68/37)  & $0.86^{+0.07}_{-0.00}$ & $< 0.04 (0.01)$ & 1.13 (41.97/37	\\
			&					 & SMC 	& $0.43^{+0.33}_{-0.38}$ & $0.11^{+0.09}_{-0.08}$ & 1.04 (38.49/37) & $0.86^{+0.07}_{-0.00}$ & $< 0.04 (0.00)$ & 1.14 (42.15/37)	\\	
\hline 
GRB\,081121  &  	43, 430			          & MW  & $-0.13^{+0.25}_{-0.28}$ & $0.10^{+0.03}_{-0.03}$ & 0.65 (3.93/6) & $0.37^{+0.13}_{-0.00}$ & $0.08^{+0.01}_{-0.02}$ & 0.95 (5.71/6)		\\
			&					 & LMC & $-0.38^{+0.50}_{-0.43}$ & $0.15^{+0.05}_{-0.05}$ & 0.29 (1.74/6) & $0.37^{+0.11}_{-0.00}$ & $0.07^{+0.01}_{-0.01}$ & 1.43 (8.55/6)	 	\\
			&					 & SMC & $-0.31^{+0.43}_{-0.41}$ & $0.11^{+0.03}_{-0.03}$ & 0.53 (3.18/6) & $0.37^{+0.12}_{-0.00}$ & $0.06^{+0.01}_{-0.01}$ & 1.48 (8.87/6)	 	 \\	
\hline 
GRB\,081203A  &  	140, 7200			 & MW  & $0.68^{+0.95}_{-1.31}$ & $0.17^{+0.07}_{-0.11}$ & 0.75 (5.27/7) & $0.66^{+0.11}_{-0.08}$ & $0.17^{+0.03}_{-0.03}$ & 0.75 (5.27/7) \\
			&					 & LMC & $-1.45^{+2.30}_{-2.23}$ & $0.29^{+0.16}_{-0.15}$ & 0.74 (5.15/7) & $0.57^{+0.02}_{-0.17}$ & $0.15^{+0.02}_{-0.03}$ &0.92 (6.42/7) \\
			&					 & SMC & $-5.92^{+4.51}_{-4.18}$ & $0.49^{+0.24}_{-0.25}$ & 0.74 (5.15/7) & $0.67^{+0.12}_{-0.07}$ & $0.12^{+0.02}_{-0.02}$ &1.16 (8.14/7)  \\	
\hline 
GRB\,081221  &  	190			          & MW  & & & & 0.89 & $> 0.5$ & \\
			&					 & LMC & & & & 0.89 & $> 0.5$ & \\
			&					 & SMC & & & & 0.89 & $> 0.5$ & \\	
\hline 
GRB\,081222  &  	1.2, 43			 & MW & $-0.45$ & $0.02$ & 1.45 (27.54/19) & $0.48$ & $< 0.03 (0.02)$ & 1.45 (27.54/19)  \\
			&					 & LMC & $-0.31^{+0.70}_{-0.71}$ & $0.15^{+0.12}_{-0.12}$ & 1.30 (24.68/19) & $0.47^{+0.12}_{-0.00}$ & $< 0.03 (0.01)$ & 1.43 (27.19/19)  \\
			&					 & SMC & $-0.01^{+0.52}_{-0.54}$ & $< 0.19 ( 0.07)$ & 1.33 (25.19/19) & $0.47^{+0.12}_{-0.00}$ & $< 0.02 (0.01)$ & 1.42 (27.03/19)  \\	
\hline 
GRB\,090102  &  	4.3, 29		& MW & $0.95^{+0.14}_{-0.35}$ & $< 0.20 (0.00)$ & 1.06 (6.33/6)& $0.36^{+0.14}_{-0.00}$ & $0.14_{-0.03}^{+0.06}$ & 1.71 (10.27/6) \\
			&				& LMC & $-0.16^{+1.00}_{-1.00}$ & $0.21^{+0.21}_{-0.19}$ & 0.67 (4.00/6) & $0.22^{+0.14}_{-0.00}$ & $0.14^{+0.02}_{-0.05}$ & 0.71 (4.27/6) \\
			&				& SMC & $0.27^{+0.54}_{-0.59}$ & $0.11^{+0.10}_{-0.09}$ & 0.54 (3.23/6) & $0.27^{+0.09}_{-0.05}$ & $0.11^{+0.03}_{-0.03}$ & 0.54 (3.23/6) \\	
\hline 
GRB\,090201   &  	434			          & MW  & & & & 0.49 & $> 0.4$ & \\
			&					 & LMC & & & & 0.49 & $> 0.4$ &  \\
			&					 & SMC & & & & 0.49 & $> 0.4$ &  \\	
\hline 
GRB\,090424 &  	120, 4300			 & MW 	& $1.35^{+0.25}_{-0.27}$ & $0.09^{+0.09}_{-0.08}$ & 0.85 (27.24/32) & $0.55^{+0.00}_{-0.06}$ & $0.35^{+0.02}_{-0.01}$ & 1.31 (41.81/32)	\\
			&					 & LMC 	& $1.32^{+0.29}_{-0.21}$ & $0.09^{+0.06}_{-0.05}$ & 0.75 (23.97/32) & $0.55$ & $0.31$ & 1.50 (47.97/32) \\
			&					 & SMC & $1.41^{+0.20}_{-0.15}$ & $0.06^{+0.04}_{-0.04}$ & 0.74 (23.80/32)	 & $0.55$ & $0.31$ & 2.15 (68.70/32)  \\	
\hline 
GRB\,090709A  &  	1.4, 14			 & MW & $-0.03^{+3.20}_{-7.64}$ & $< 5.88$ & 1.84 (9.18/5) & $0.35^{+0.09}_{-0.05}$ & $0.83^{+0.16}_{-0.16}$ & 1.84 (9.19/5) \\
			&					 & LMC & $0.24^{+2.93}_{-7.48}$ & $< 5.02$ & 1.84 (9.19/5) & $0.44^{+0.00}_{-0.14}$ & $0.75^{+0.18}_{-0.11}$ & 1.84 (9.20/5) \\
			&					 & SMC & $1.52$ & $0.45$ & 1.87 (9.36/5) & $0.44$ & $0.77$ & 1.90 (9.49/5) \\	
\hline 
GRB\,090715B  &  	8.6, 430			 & MW & $1.40^{+0.12}_{-0.12}$ & $0.18^{+0.07}_{-0.08}$ & 0.67 (31.94/48)	& 0.63 & 0.00 & 2.24 (107.63/48) \\
			&					 & LMC & $0.11^{+0.38}_{-0.44}$ & $0.17^{+0.07}_{-0.06}$ & 0.66 (31.84/48)	 & $0.45^{+0.00}_{-0.18}$ & $0.11^{+0.01}_{-0.03}$ & 0.69 (33.20/48)	\\
			&					 & SMC & $-1.09^{+0.85}_{-0.91}$ & $0.18^{+0.08}_{-0.07}$ & 0.66 (31.78/48) & $0.45^{+0.00}_{-0.18}$ & $0.06^{+0.01}_{-0.02}$ & 0.79 (38.07/48) \\	
\hline 
GRB\,090812  &  	1.4, 290			 & MW & $1.24^{+0.26}_{-0.31}$ & $< 0.16 (0.03)$ & 1.52 (12.15/8) & $0.52$ & $0.16$ & 2.65 (21.20/8) \\
			&					 & LMC & $0.54^{+0.82}_{-0.89}$ & $< 0.33 (0.12)$ & 1.31 (10.51/8) &  $0.51^{+0.01}_{-0.12}$ & $0.12_{-0.03}^{+0.04}$ & 1.31 (10.52/8)Ê\\
			&					 & SMC & $0.58^{+0.85}_{-1.01}$ & $< 0.24 (0.08)$ & 1.38 (11.08/8) &  $0.52^{+0.00}_{-0.13}$ & $0.08_{-0.02}^{+0.03}$ & 1.39 (11.09/8) \\	
\hline
GRB\,090926B &  				          & MW 		\\
			&					 & LMC 	 	\\
			&					 & SMC 	 	 \\	
\hline
\end{tabular}
\end{small}
\end{table*}

\begin{table*}
\contcaption{}
\begin{small}
\begin{tabular}{lclccclcccccc}
\hline 
GRB\,091018 &  	86, 230	& MW & $0.57^{+0.01}_{-0.01}$ & $< 0.01 (0.00)$ & 1.15 (93.54/81) & $0.57^{+0.01}_{-0.02}$ & $< 0.01 (0.00)$ & 1.16 (93.64/81		\\
			&			& LMC & $0.57^{+0.01}_{-0.01}$ & $< 0.01 (0.00)$ & 1.15 (93.51/81) & $0.57^{+0.01}_{-0.02}$ & $< 0.01 (0.00)$ & 1.16 (93.64/81)	 	\\
			&			& SMC & $0.57^{+0.01}_{-0.06}$ & $< 0.02 (0.00)$ & 1.15 (93.51/81) &	$0.57^{+0.01}_{-0.06}$ & $< 0.02 (0.00)$ & 1.16 (93.64/81) 	 \\	
\hline 
GRB\,091020 &  	12, 720	& MW 	& $2.02^{+0.15}_{-0.16}$ & $0.10^{+0.02}_{-0.03}$ & 1.40 (23.79/17) & $0.66$ & $0.29$ & 13.99 (237.76/17) \\
			&			& LMC 	& $0.23^{+0.65}_{-0.64}$ & $0.35^{+0.10}_{-0.10}$ & 1.39 (23.61/17) & $0.55^{+0.11}_{-0.00}$ & $0.30^{+0.01}_{-0.03}$ & 1.42 (24.12/17) \\
			&			& SMC 	& $2.38$ & $0.00$ & 2.70 (45.98/17) & $0.66$ & $0.18$ & 4.45 (75.62/17)  \\	
\hline 
GRB\,091127 &  	86, 240	& MW 	& $0.23^{+0.02}_{-0.02}$ & $0.04^{+0.01}_{-0.01}$ & 0.93 (367.11/396) 	& $0.22^{+0.02}_{-0.01}$ & $0.04^{+0.01}_{-0.01}$ & 0.93 (367.23/396) \\
			&			& LMC 	& $0.23^{+0.03}_{-0.03}$ & $0.04^{+0.01}_{-0.01}$ & 0.95 (374.56/396) 	 & $0.22^{+0.04}_{-0.02}$ & $0.04^{+0.01}_{-0.01}$ & 0.95 (375.96/396) \\
			&			& SMC 	& $0.28^{+0.03}_{-0.02}$ & $0.02^{+0.01}_{-0.01}$ & 0.98 (387.70/396) 	 & $0.29^{+0.00}_{-0.04}$ & $0.02^{+0.01}_{-0.01}$ & 0.98 (387.95/396) \\	
\hline  
GRB\,091208B  &  		580, 1200		          & MW & $1.24^{+0.53}_{-0.64}$ & $< 0.19 (0.07)$ & 1.08 (6.51/6)	&  $1.07^{+0.00}_{-0.21}$ & $0.09_{-0.03}^{+0.07}$ & 1.11 (6.66/6)	 \\
			&					 & LMC & $1.08^{+0.68}_{-0.88}$ & $< 0.29 (0.10)$ & 1.08 (6.48/6) &  $1.07^{+0.00}_{-0.21}$ & $0.10_{-0.03}^{+0.08}$ & 1.08 (6.48/6)	 \\
			&					 & SMC & $0.94^{+0.83}_{-1.19}$ & $< 0.43 (0.14)$ & 1.09 (6.54/6) & $0.94^{+0.13}_{-0.08}$ & $0.14_{-0.07}^{+0.06}$ & 1.09 (6.54/6)	\\	
\hline
GRB\,100615A &  	24			          & MW & & & & 0.76 & $> 2.8$ &  \\
			&					 & LMC & & & & 0.76 & $> 2.8$ & \\
			&					 & SMC & & & & 0.76 & $> 3.2$ & \\	
\hline
GRB\,100621A &  	100, 140	& MW 	& $1.07^{+0.70}_{-0.71}$ & $1.16^{+0.26}_{-0.26}$ & 1.77 (7.08/4) 	& $1.02^{+0.01}_{-0.12}$ & $1.17^{+0.12}_{-0.04}$ & 1.77 (7.09/4) \\
			&			& LMC 	& $1.19^{+0.64}_{-0.65}$ & $1.04^{+0.22}_{-0.21}$ & 1.84 (7.37/4) & $1.03^{+0.00}_{-0.25}$ & $1.10^{+0.11}_{-0.03}$ & 1.88 (7.52/4	\\
			&			& SMC 	& $1.80$ & $0.87$ & 2.52 (10.09/4) 	& $1.03$ & $1.16$ & 3.49 (13.94/4) \\	
\hline 
GRB\,100728B &  	2.2, 140  	& MW  & $0.45^{+0.50}_{-0.48}$ & $< 0.23 (0.08)$ & 0.45 (2.70/6) 	& $0.47^{+0.28}_{-0.07}$ & $< 0.14 (0.07)$ & 0.45 (2.70/6) 	\\
			&			& LMC & $0.29^{+0.66}_{-0.81}$ & $< 0.35 (0.10)$ & 0.45 (2.68/6) 	& $0.40^{+0.35}_{-0.00}$ & $< 0.14 (0.08)$ & 0.46 (2.74/6) \\
			&			& SMC & $0.70^{+0.31}_{-0.46}$ & $< 0.17 (0.02)$ & 0.60 (3.58/6) 	& $0.70^{+0.05}_{-0.30}$ & $< 0.11 (0.02)$ & 0.60 (3.58/6) \\	
\hline
GRB\,110205A &  	360, 580	& MW & $1.23$ & $0.00$ & 1.72 (139.47/81) & $0.72$ & $0.12$ & 9.00 (729.07/81)	 \\
			&			& LMC & $0.55_{-0.12}^{+0.10}$ & $0.11_{-0.01}^{+0.02}$ & 1.18 (95.39/81) & $0.55_{-0.01}^{+0.14}$ & $0.11_{-0.02}^{+0.01}$ & 1.18 (95.40/81) \\
			&			& SMC & $0.78$ & $0.04$ & 1.28 (104.01/81) 	& $0.72$ & $0.05$ & 1.29 (104.57/81) \\	
\hline
GRB\,110503A &  	4.3, 720	& MW  & $1.24_{-0.45}^{+0.40}$ & $< 0.23 (0.02)$ & 1.27 (2.55/2) 	& $0.49$ & $0.17$ & 4.64 (9.27/2) \\
			&			& LMC & $1.29_{-0.87}^{+0.36}$ & $< 0.24 (0.00)$ & 1.31 (2.63/2)  & $0.49_{-0.10}^{+0.00}$ & $0.10_{-0.05}^{+0.07}$ & 2.21 (4.42/2) \\
			&			& SMC & $1.28_{-0.74}^{+0.36}$ & $< 0.14 (0.00)$ & 1.31 (2.63/2)	& $0.49$ & $0.06$ & 2.45 (4.90/2)  \\	
\hline
\end{tabular}
\end{small}
\end{table*}

\bsp

\label{lastpage}

\end{document}